\pdfoutput=1 

\documentclass[11pt]{article}

\usepackage{booktabs}
\usepackage{hyperref}
\usepackage{caption}
\usepackage{subcaption}

\usepackage{multirow}

\usepackage{xspace}
\usepackage{setspace}
\usepackage{enumerate}

\usepackage{censor}

\usepackage{listings}

\usepackage{microtype}

\usepackage[utf8]{inputenc}

\DisableLigatures{}

\usepackage{xcolor}
\definecolor{codegreen}{rgb}{0.25,0.5,0.35}
\definecolor{codegray}{rgb}{0.5,0.5,0.5}
\definecolor{codepurple}{rgb}{0.6,0,0}
\definecolor{backcolour}{rgb}{0.95,0.95,0.92}
\definecolor{colorstring}{rgb}{0.5,0,0.35}
\definecolor{rltred}{rgb}{0.5,0,0}
\definecolor{rltgreen}{rgb}{0,0.5,0}
\definecolor{rltblue}{rgb}{0,0,0.5}
\definecolor{DarkGreen}{rgb}{0.00,0.60,0.00}
\definecolor{ScarletRed}{rgb}{0.80,0.00,0.00}
\definecolor{blizzardblue}{rgb}{0.67, 0.9, 0.93}
\definecolor{green-yellow}{rgb}{0.68, 1.0, 0.18}
\definecolor{dkgreen}{rgb}{0,0.6,0}
\definecolor{gray}{rgb}{0.5,0.5,0.5}
\definecolor{mauve}{rgb}{0.58,0,0.82}
\definecolor{lightgrey}{rgb}{0.90,0.90,0.90}
\definecolor{grey}{gray}{0.75}
\definecolor{light-gray}{gray}{0.80}

\lstdefinestyle{mystyle}{
	backgroundcolor=\color{backcolour},   
	commentstyle=\color{codegreen},
	keywordstyle=\color{colorstring}\bfseries,
	numberstyle=\ttfamily\color{codegray},
	stringstyle=\color{codepurple},
            basicstyle={\scriptsize\ttfamily},
	breakatwhitespace=false,         
	breaklines=true,                 
	captionpos=b,                    
	keepspaces=true,                 
	numbers=left,                    
	numbersep=2pt,                  
	showspaces=false,                
	showstringspaces=false,
	showtabs=false,                  
	tabsize=2
}
\usepackage{geometry}
\newcommand{\bBOXRT}{{\sc bBOXRT}\xspace}
\newcommand{\evo}{{\sc EvoMaster}\xspace}

\newcommand{\morest}{{\sc Morest}\xspace}
\newcommand{\Pythia}{{\sc Pythia}\xspace}
\newcommand{\RESTest}{{\sc RESTest}\xspace}
\newcommand{\RestTest}{{\sc RESTest}\xspace}
\newcommand{\restest}{{\sc RESTest}\xspace}
\newcommand{\RestCT}{{\sc RestCT}\xspace}
\newcommand{\RESTler}{{\sc RESTler}\xspace}
\newcommand{\Restler}{{\sc RESTler}\xspace}
\newcommand{\RestTestGen}{{\sc RestTestGen}\xspace}
\newcommand{\Schemathesis}{{\sc Schemathesis}\xspace}
\newcommand{\restats}{{\sc Restats}\xspace}

\newcommand{\ExVivoMicroTest}{{\sc ExVivoMicroTest}\xspace}
\newcommand{\HsuanFuzz}{{\sc HsuanFuzz}\xspace}
\newcommand{\QuickRest}{{\sc QuickRest}\xspace}

\newcommand{\etal}{{\emph{et al.}}\xspace}
\newcommand{\eg}{{e.g.,}\xspace}
\newcommand{\ie}{{i.e.,}\xspace}

\newcommand{\mCovSchema}{\cite{ \papersecond,\paperseventh,\papereleventh,\papertwentysixth,\paperfortyfirst,\paperfiftysecond,\papersixtysecond,\papersixtythird,\papereightyfirst }\xspace}

\newcommand{\mCovSpec}{\cite{ \paperthirtyfirst,\paperfortyfirst,\paperfortysixth,\papersixtieth,\papersixtyfirst,\papersixtythird,\papersixtyeighth,\paperseventyninth,\papereightyeighth,\paperninetieth,\paperninetysecond,\paperninetysixth }\xspace}
\newcommand{\mFaultFivexx}{\cite{ \paperfirst,\papersecond,\paperthird,\paperfourth,\paperfifth,\papersixth,\paperseventh,\papertenth,\papereleventh,\papersixteenth,\paperseventeenth,\papertwentyfirst,\papertwentythird,\papertwentyfifth,\papertwentysixth,\paperthirtyeighth,\paperthirtyninth,\paperfortyfirst,\paperfortythird,\paperfortyfifth,\paperfiftieth,\paperfiftysecond,\paperfiftyfifth,\papersixtythird,\papersixtyfourth,\papersixtysixth,\paperseventieth,\paperseventyfirst,\papereightyfirst,\paperninetysecond }\xspace}
\newcommand{\mFaultSchema}{\cite{ \paperfirst,\paperthird,\paperfifth,\papersixth,\paperseventh,\papertenth,\paperseventeenth,\papertwentysixth,\paperthirtyeighth,\paperthirtyninth,\paperfortyfifth,\paperfiftieth,\paperseventieth,\paperseventyfirst }\xspace}
\newcommand{\mFaultRule}{\cite{ \paperfifth,\paperfourteenth,\papersixteenth,\papertwentieth,\papertwentyseventh,\papertwentyeighth,\papertwentyninth,\paperfortyfirst,\paperfortyninth,\paperfiftyfifth,\papersixtyfirst,\papersixtysixth,\papersixtyeighth,\paperninetysixth }\xspace}

\newcommand{\preOfBB}{73\%\xspace}

\newcommand{\paperfirst}{corradiniautomated2022}
\newcommand{\papersecond}{ban2021automated}
\newcommand{\paperthird}{zhang2021resource}
\newcommand{\paperfourth}{sahin2021discrete}
\newcommand{\paperfifth}{laranjeiro2021black}
\newcommand{\papersixth}{arcuri2020blackbox}
\newcommand{\paperseventh}{martin2021black}
\newcommand{\papereighth}{mirabella2021deep}
\newcommand{\paperninth}{corradini2021empirical}
\newcommand{\papertenth}{stallenberg2021improving}
\newcommand{\papereleventh}{martinLopez2021Restest}
\newcommand{\papertwelfth}{vassiliou2020simple}
\newcommand{\paperthirteenth}{martin2020ai}
\newcommand{\paperfourteenth}{atlidakis2020checking}
\newcommand{\paperfifteenth}{khortiuk2020increasing}
\newcommand{\papersixteenth}{karlsson2020QuickREST}
\newcommand{\paperseventeenth}{viglianisi2020resttestgen}
\newcommand{\papereighteenth}{Godefroid2020Restlerdata}
\newcommand{\papernineteenth}{zhao2020semantic}
\newcommand{\papertwentieth}{vu2019model}
\newcommand{\papertwentyfirst}{restlerICSE2019}
\newcommand{\papertwentysecond}{bhagya2019generating}
\newcommand{\papertwentythird}{arcuri2019restful}
\newcommand{\papertwentyfourth}{martin2019test}
\newcommand{\papertwentyfifth}{zhang2019resource}
\newcommand{\papertwentysixth}{eddouibi2018automatic}
\newcommand{\papertwentyseventh}{segura2017metamorphic}
\newcommand{\papertwentyeighth}{katt2018quantitative}
\newcommand{\papertwentyninth}{fertig2015model}
\newcommand{\paperthirtieth}{lamela2013towards}
\newcommand{\paperthirtyfirst}{chakrabarti2010connectedness}
\newcommand{\paperthirtysecond}{chakrabarti2009test}
\newcommand{\paperthirtythird}{corradini2021restats}
\newcommand{\paperthirtyfourth}{takeda2019applying}
\newcommand{\paperthirtyfifth}{wenhui2017study}
\newcommand{\paperthirtysixth}{godefroid2020differential}
\newcommand{\paperthirtyseventh}{masood2015static}
\newcommand{\paperthirtyeighth}{zhang2021adaptive}
\newcommand{\paperthirtyninth}{arcuri2020handling}
\newcommand{\paperfortieth}{zhang2022open}
\newcommand{\paperfortyfirst}{valenzuela2022arte}
\newcommand{\paperfortysecond}{bucaille2020openapi}
\newcommand{\paperfortythird}{wu2022icse}
\newcommand{\paperfortyfourth}{Kim2022Rest}
\newcommand{\paperfortyfifth}{zhang2021enhancing}
\newcommand{\paperfortysixth}{pinheiro2013model}
\newcommand{\paperfortyeighth}{lin2022forest}
\newcommand{\paperfortyninth}{barabanov2022automatic}
\newcommand{\paperfiftieth}{arcuri2021tt}
\newcommand{\paperfiftyfirst}{marculescu2022faults}
\newcommand{\paperfiftysecond}{tsai2021rest}
\newcommand{\paperfiftythird}{martinicse}
\newcommand{\paperfiftyfourth}{olsthoorn2022more}
\newcommand{\paperfiftyfifth}{hatfield2022deriving}
\newcommand{\paperfiftysixth}{martin2021specification}
\newcommand{\paperfiftyseventh}{vassiliou2021quality}
\newcommand{\paperfiftyeighth}{vassiliou2022solving}
\newcommand{\papersixtieth}{benac2014jsongen}
\newcommand{\papersixtyfirst}{cheh2021analyzing}
\newcommand{\papersixtysecond}{yamamoto2021efficient}
\newcommand{\papersixtythird}{liu2022icse}
\newcommand{\papersixtyfourth}{atlidakis2020pythia}
\newcommand{\papersixtyfifth}{chen2021bootstrapping}
\newcommand{\papersixtysixth}{barlas2022exploiting}
\newcommand{\papersixtyseventh}{yamamoto2020call}
\newcommand{\papersixtyeighth}{zhahazard}
\newcommand{\papersixtyninth}{mahmood2022framework}
\newcommand{\paperseventieth}{arcuri2020testability}
\newcommand{\paperseventyfirst}{arcuri2019sql}
\newcommand{\paperseventysecond}{karlsson2019exploratory}
\newcommand{\paperseventythird}{arcuri2021evomaster}
\newcommand{\paperseventyfourth}{wolde2021rest}
\newcommand{\paperseventyfifth}{bondel2021design}
\newcommand{\paperseventysixth}{soni2020mockrest}
\newcommand{\paperseventyseventh}{idris2021development}
\newcommand{\paperseventyeighth}{manikantan2022software}
\newcommand{\paperseventyninth}{liu2018cpn}
\newcommand{\papereightieth}{vu2018automation}
\newcommand{\papereightyfirst}{arcuri2017restful}
\newcommand{\papereightysecond}{sotiriadis2017unit}
\newcommand{\papereightythird}{ballesterostool}
\newcommand{\papereightyfourth}{besson2011towards}
\newcommand{\papereightyfifth}{liu2017optimized}
\newcommand{\papereightysixth}{navas2014rest}

\newcommand{\papereightyeighth}{gazzola2022exvivomicrotest}
\newcommand{\paperninetieth}{luo2019verification}
\newcommand{\paperninetyfirst}{sotiriadis2018testing}
\newcommand{\paperninetysecond}{restest2020}
\newcommand{\paperninetyfourth}{vu2018verification}
\newcommand{\paperninetysixth}{mai2020metamorphic}
\newcommand{\paperninetyseventh}{martin2019catalogue}
\newcommand{\paperninetyeighth}{js2022}

\newcommand{\rqAYear}{\textbf{RQ1}: How many papers were published per year?}
\newcommand{\rqAVenues}{\textbf{RQ2}: In which venues were the papers published?}
\newcommand{\rqAContri}{\textbf{RQ3}: What contributions to testing RESTful APIs have been made by the papers in this area, and what are their frequencies?}

\newcommand{\rqBMetric}{\textbf{RQ4}: What metrics are used to evaluate the effectiveness of the testing?}
\newcommand{\rqBTechni}{\textbf{RQ5}: What techniques are used for automatically testing RESTful APIs?}
\newcommand{\rqBTestKind}{\textbf{RQ6}: What kind of testing has been automated for RESTful APIs?}
\newcommand{\rqBCaseStudy}{\textbf{RQ7}: What kind of artifacts are used for conducting empirical evaluations?}

\newcommand{\rqCOpenSou}{\textbf{RQ8}: Which research tools are open-source?}
\newcommand{\rqCNonResTool}{\textbf{RQ9}: Which non-research tools are used/compared?}
\newcommand{\rqCFeatures}{\textbf{RQ10}:  Which features are supported by the research prototypes?}

\newcommand{\rqDAdresChal}{\textbf{RQ11}: Which research challenges are addressed?}
\newcommand{\rqDOpenChal}{\textbf{RQ12}: Which open research challenges are identified?}

\newcommand{\totalPapers}{{92}\xspace} 
\newcommand{\searchPapers}{{40}\xspace} 
\newcommand{\snowballingPapers}{{52}\xspace} 
\newcommand{\papersAfterMarch}{{16}\xspace} 
\newcommand{\papersFromOtherDatabases}{{15}\xspace} 
 
\newcommand{\otherDatabases}{{\textit{arXiv}, \textit{Penn State}, \textit{KSIResearch}, \textit{Open Journals}, \textit{unam.mx}, \textit{Semantic Scholar}, \textit{sistedes.es} and \textit{Macrothink Institute}}\xspace} 

\newcommand{\NumberOfPapersByUs}{{16}\xspace} 

\newcommand{\papersUsingEmb}{{\cite{\paperthird,\paperfourth,\papersixth,\papertenth,\papertwentythird,\papertwentyfourth,\papertwentyfifth,\paperthirtyeighth,\paperthirtyninth,\paperfortieth,\paperfortyfifth,\paperfiftieth,\paperfiftyfirst,\papersixtyseventh,\paperseventieth,\paperseventyfirst,\paperninetyeighth}}\xspace} 
\newcommand{\numberOfPapersUsingEmb}{{17}\xspace}

\newcommand{\paperstwentyeighteen}{{8}\xspace} 
\newcommand{\paperstwentytwentyone}{{24}\xspace}

\newcommand{\numberOfVenues}{{63}\xspace} 
\newcommand{\numberOfJournals}{{14}\xspace} 
\newcommand{\numberOfConferences}{{48}\xspace} 
\newcommand{\numberOfOpenAccessRepos}{{1}\xspace} 

\newcommand{\journalPapers}{{20}\xspace} 
\newcommand{\conferencePapers}{{66}\xspace} 
\newcommand{\openAccessRepoPapers}{{6}\xspace} 

\newcommand{\otherConferences}{{43}\xspace}

\newcommand{\topVenueNumber}{{7}\xspace} 

\newcommand{\topJournal}{{TOSEM}\xspace}

\newcommand{\firstYear}{{2009}\xspace} 
\newcommand{\lastYear}{{2022}\xspace}

\newcommand{\EA}{{Empirical study on various automated approaches}\xspace} 
 
\newcommand{\EAPercent}{{3\%}\xspace} 

\newcommand{\NAlower}{{new automated approach and its extension}\xspace} 
\newcommand{\NA}{{New automated approach and its extension}\xspace} 
\newcommand{\NANumber}{{61}\xspace} 
\newcommand{\NAPercent}{{66\%}\xspace} 

\newcommand{\NPlower}{{new analysis approach and potential solution}\xspace} 
\newcommand{\NP}{{New analysis approach and potential solution}\xspace} 
 
\newcommand{\NPPercent}{{18\%}\xspace}

\newcommand{\PI}{{Proposal or idea}\xspace} 
\newcommand{\PINumber}{{4}\xspace}

\newcommand{\TD}{{Tool implementation or demostration}\xspace}

\newcommand{\NumberOfContributionTypes}{{5}\xspace} 

\newcommand{\testingTypes}{{8}\xspace} 
\newcommand{\System}{{72}\xspace}

\newcommand{\Regression}{{2}\xspace}

\newcommand{\LocalOpenSource}{{30\%}\xspace} 

\newcommand{\Internet}{{28\%}\xspace} 

\newcommand{\LocalArtificial}{{16\%}\xspace} 

\newcommand{\NoCaseStudies}{{14\%}\xspace} 

\newcommand{\LocalClosedSource}{{7\%}\xspace} 

\newcommand{\LocalOpenSourceandClosedSource}{{4\%}\xspace}

\newcommand{\NumberOfOpenSourceTools}{{16}\xspace} 
\newcommand{\NumberOfPapersWithOpenSourceTools}{{44}\xspace} 

\newcommand{\NumberOfPapersUsingEvoMaster}{{19}\xspace} 
 
\newcommand{\NumberOfPapersUsingRestler}{{8}\xspace}

\newcommand{\MostUsedProgrammingLanguageByOpenSourceTool}{{Java}\xspace} 
\newcommand{\NumberOfMostUsedProgrammingLanguageByOpenSourceTool}{{6}\xspace} 
\newcommand{\SecondMostUsedProgrammingLanguageByOpenSourceTool}{{Python}\xspace}

\newcommand{\NumberOfTools}{{16}\xspace} 

\newcommand{\NumberOfNonResearchTools}{{20}\xspace} 

\newcommand{\restassuredNumber}{{12}\xspace}

\newcommand{\burpsuiteNumber}{{2}\xspace}

\newcommand{\tntfuzzerNumber}{{2}\xspace}

\newcommand{\NumberOfToolsSupportingAuth}{{8}\xspace} 
\newcommand{\NumberOfToolsSupportingOpenAPI}{{12}\xspace}

\newcommand{\handlingresourcedependenciesNumber}{{8}\xspace} 
\newcommand{\inferringinterparameterdependenciesNumber}{{6}\xspace}

\newcommand{\PapersWithOpenChallenges}{{73}\xspace}

\newcommand{\toolsupport}{{25}\xspace} 

\newcommand{\havingmorerestcasestudies}{{13}\xspace} 

\newcommand{\securitytesting}{{10}\xspace}

\newcommand{\handlingexternalservices}{{4}\xspace}

\newcommand{\RQTenCategoriesNumber}{{five}\xspace}

\usepackage{pgfplots} 

\definecolor{ForestGreen}{RGB}{34,139,34}
\definecolor{asparagus}{rgb}{0.53, 0.66, 0.42}

\usepackage{boxedminipage}
\newenvironment{result}%
{\smallskip
	\noindent
	\let\emph=\textbf
	\begin{boxedminipage}{\columnwidth}\begin{center}\em}%
		{\end{center}\end{boxedminipage}%
}

\newboolean{showcomments}
\setboolean{showcomments}{true} % comment this line to deactivate comments

\ifthenelse{\boolean{showcomments}}{
	\newcommand{\nbc}[3]{
		{\colorbox{#3}{\bfseries\sffamily\scriptsize\textcolor{white}{#1}}}
		{\textcolor{#3}{\sf\small$\langle$\textit{#2}$\rangle$}}}
	
}{
	\newcommand{\nbc}[3]{}

}

\newcommand{\paragraphfourth}[1]{\paragraph{\textbf{#1}}\mbox{\\}}
\newcommand{\paragraphfifith}[1]{\textit{\textbf{#1.}}}

\title{Testing RESTful APIs: A Survey}
\author{Amid Golmohammadi, Man Zhang, Andrea Arcuri\\
Kristiania University College and Oslo Metropolitan University, Norway}
\date{}

\begin{document}

\maketitle

\begin{abstract}
In industry, RESTful APIs are widely used to build modern Cloud Applications. 
Testing them is challenging, because not only they rely on network communications, but also they deal with external services like databases.
Therefore, there has been a large amount of research sprout in recent years on how to automatically verify this kind of web services.
In this paper, we review the current state-of-the-art on testing RESTful APIs, based on an analysis of \totalPapers scientific articles. 
This review categories and summarizes the existing scientific work on this topic, and discusses the current challenges in the verification of RESTful APIs.
\end{abstract}


%%%%%%%%%%%%%%%%%%%%%%%%%%%%%%%%%%%%%%%%%%%%%%%%%%%%%%%%%%%%%%%%%%%%%%%%%%%%
\section{Introduction}
\label{sec:intro}
When building the backends for web and enterprise applications, using RESTful APIs~\cite{fielding2000architectural} is a common choice, especially in microservice architectures~\cite{newman2015building}.
Many companies use this kind of systems for their backends, like for example  Netflix, Uber, Airbnb, eBay, Amazon, Twitter, Nike, etc~\cite{rajesh2016spring}.

Not only RESTful APIs are used to build the backends of enterprise applications, but also there are many of this kind of APIs directly available on internet, providing all different kinds of functionalities.
For example,
\emph{ProgrammableWeb}\footnote{https://www.programmableweb.com/api-research} 
currently lists more than 24 thousand Web APIs, where RESTful ones are the most common (other kinds are for example the old SOAP~\cite{curbera2002unraveling} and the more recent GraphQL~\cite{GraphQLFoundation}).
Several companies provide APIs to their services on internet using REST,
such as for example
Google\footnote{https://developers.google.com/drive/v2/reference/},
Amazon\footnote{http://docs.aws.amazon.com/AmazonS3/latest/API/Welcome.html},
Twitter\footnote{https://dev.twitter.com/rest/public},
Reddit\footnote{https://www.reddit.com/dev/api/},
and
LinkedIn\footnote{https://docs.microsoft.com/en-us/linkedin/}.

However, verifying the correctness of Web APIs is quite challenging~\cite{canfora2009service,bozkurt2013testing}.
Not only a tester needs to create network messages (e.g., using HTTP over TCP) toward the API, but typically there is also the need to setup the right data into the databases  and possibly mock interactions with other external services~\cite{arcuri2018experience}.
Due to their wide use in industry, it is hence not surprising that there has been a lot of attention from the research community on the developing of novel techniques to test this kind of applications in recent years. 
Therefore, to help carrying out further research endeavors on this topic, in this paper we survey and categorize the current state-of-the-art of the scientific literature on testing RESTful APIs.
In particular, in this paper we aim at answering the following 12 research questions from four perspectives based on a selection of \totalPapers scientific articles:

\begin{itemize}
	\item Publication status in testing RESTful APIs
	\begin{itemize}
		\item \rqAYear
		\item \rqAVenues
		\item \rqAContri
	\end{itemize}
	\item Existing approaches in supporting automated testing of RESTful APIs
	\begin{itemize}
		\item \rqBMetric
		\item \rqBTechni
		\item \rqBTestKind
		\item \rqBCaseStudy
	\end{itemize}
	\item Available Tools for Testing RESTful APIs
	\begin{itemize}
		\item \rqCOpenSou
		\item \rqCNonResTool
		\item \rqCFeatures
	\end{itemize}
	\item Addressed and Open Challenges
	\begin{itemize}
		\item \rqDAdresChal
		\item \rqDOpenChal
	\end{itemize}
\end{itemize}

The paper is organized as follows.
Section~\ref{sec:background} discusses important background information, needed to better understand the rest of the paper.
Related work is discussed then in Section~\ref{sec:related}.
How the \totalPapers articles were selected is described in Section~\ref{sec:method}.
Our research questions are answered in
Section~\ref{sec:survey} (publication status, RQ1-3),
Section~\ref{sec:approaches} (approaches, RQ4-7),
Section~\ref{sec:tools} (tools, RQ8-10),
and
Section~\ref{sec:challenges} (challenges, RQ11-12).
Threats to validity are discussed in Section~\ref{sec:threats}.
Finally, Section~\ref{sec:conclusions} concludes the paper.

%%%%%%%%%%%%%%%%%%%%%%%%%%%%%%%%%%%%%%%%%%%%%%%%%%%%%%%%%%%%%%%%%%%%%%%%%%%%
\section{Background}
\label{sec:background}

%----------------------------------------------------------
\subsection{Hypertext Transfer Protocol and REST}
\label{subsec:http}
HTTP (Hypertext Transfer Protocol) is an application-layer protocol for hypermedia information systems that are distributed and collaborative over a computer network.
Through the extension of its request methods, error codes, and headers, this generic and stateless protocol can be used for many tasks other than hypertext, such as name servers and distributed object management systems~\cite{fielding1999hypertext}.

REST (Representational State Transfer) is an architectural style which was introduced by Fielding in 2000~\cite{fielding2000architectural} that is followed by many modern web APIs.
REST is not a protocol, as it just defines a set of guidelines for designing APIs for accessing and manipulating resources using HTTP over the network. 

To reduce interaction latency, enforce security, and encapsulate legacy systems, REST stresses scalability of component interactions, generality of interfaces, autonomous deployment of components, and intermediary components. 
REST does not enforce any rules on how it should be implemented at the lower level.
Instead, it defines a set of high-level design constraints including separation of concerns related to client and server, statelessness, the ability to cache data, having a uniform interface between components, multilayeredness and code-on-demand (this latter is optional).
It encourages users to come up with their own solutions as it is neither a protocol, nor a standard. 
This means that it can be implemented in a variety of ways and there is no enforcement to adopt any specific design pattern. 
For instance, REST does not force the application to embrace \textit{SOLID}~\cite{solidPrinciples} principles such as \textit{Dependency Inversion}.
Despite it is better to adhere to those kinds of design principles to have more sustainable applications, none of the REST constraints get violated in case a high level module (\eg the code which defines API endpoints) is tightly coupled to a low level module (\eg a module to connect to an SQL database).
A REST API can be implemented with different programming languages (\eg Python\footnote{https://www.python.org/}, C\#\footnote{https://docs.microsoft.com/en-us/dotnet/csharp/}) and frameworks (\eg Express\footnote{https://expressjs.com/} and Ruby on Rails\footnote{https://rubyonrails.org/}).

In a RESTful API, any piece of information that we can think of can be referred to as a resource. 
A document or image, for example, can be a REST resource, as can a temporal service, a collection of other resources, or a non-virtual object (\eg an employee)~\cite{resttutorial}. 
A RESTful API sends a \emph{representation} of the resource's state to the caller when a client request is made.
JSON (Javascript Object Notation), HTML, and plain text are among common data formats that can be sent via HTTP.
Among them, JSON is the most widely used file format~\cite{neumann2018analysis} because, contrary to its name, it is language-independent and understandable by both humans and machines. 

In order to specify the desired action to be taken for a certain resource, HTTP specifies a number of request methods. 
These request methods are commonly referred to as HTTP verbs. 
Multiple actions can be done on a resource by taking advantage of HTTP verbs such as \texttt{GET} and \texttt{POST}. 
For example, a list of employees can be retrieved by invoking the URL for employees (\eg ``\texttt{/employees}" which is appended to the base URL of the API) with \texttt{GET}.
As another example, a single employee can be fetched by calling its URL (\eg ``\texttt{/employees/42}'', where 42 is the identifier of the intended employee).
Similarly, a new employee can be added by invoking the same URL for retrieving the list of employees with \texttt{POST} verb and including the data of the new employee in the body payload of the HTTP request.
There are three other HTTP verbs that are used frequently within RESTful APIs, including \texttt{PUT}, \texttt{PATCH} and \texttt{DELETE}.
These verbs are utilized for updating by replacing, updating by modification and deleting a resource, respectively.

%----------------------------------------------------------
\subsection{OpenAPI}
\label{subsec:openapi}

The OpenAPI Specification (OAS)\footnote{https://swagger.io/specification/} defines a common, language-independent interface to RESTful APIs~\cite{Swagger}. 
It enables both humans and machines to learn about and comprehend the capabilities of the service without having access to the service's source code or network traffic analysis.
A user can comprehend and use a remote service with minimal implementation logic when it is properly defined.
An OpenAPI specification can then be utilized by automated tools for creating user-friendly documentation (e.g., an interactive web page) to display the API, tools for creating code to generate server stubs and client libraries in different programming languages, tools for testing, and many more use cases~\cite{openapispec}.

The OAS defines an OpenAPI document as an object that may be expressed in either JSON or YAML. 
Primitive data types in the OAS are based on the types supported by JSON specification~\cite{jsonspecwright}. 
Figure~\ref{fig:openapi} displays a sample OpenAPI document in JSON format. 
It gives information to the client on how to invoke endpoints in this API, such as which URIs and which HTTP verbs must be used to invoke that specific endpoint, and also what parameters might be needed and their data types. 
In this example (Figure~\ref{fig:openapi}), there are only two simple \texttt{GET} endpoints: the first one's URL is \texttt{``/employee''}, which does not take any input parameters.
The second one has the same URL, in addition to an input parameter of type \texttt{integer}.

\begin{figure}[t]
	\begin{lstlisting}[style=mystyle,language=java,label=lst:openapi]
		{
			"openapi": "3.0.1",
			"info": {
				"title": "Employees API",
				"version": "v1"
			},
			"paths": {
				"/employee/{id}": {
					"get": {
						"tags": [
						"Employee"
						],
						"parameters": [
						{
							"name": "id",
							"in": "path",
							"required": true,
							"schema": {
								"type": "integer",
								"format": "int32"
							}
						}
						],
						"responses": {
							"200": {..},
							"400": {..},
							"401": {..}
						}
					}
				},
				"/employee": {
					"get": {
						"tags": [
						"Employee"
						],
						"responses": {
							"200": {..},
							"400": {..},
							"401": {..}
						}
					}
				}
			},
			"components": {..}
		}
	\end{lstlisting}
	\caption{An example of OpenAPI 3.0 schema}
	\label{fig:openapi}
\end{figure}

OpenAPI is one of the widely used techniques to define the schema of a REST API.
There are other languages which can be used to describe APIs, such as RAML\footnote{https://raml.org/} and APIBluePrint\footnote{https://apiblueprint.org/}. However, they are not as commonly used as OpenAPI, especially for the purpose of \textit{fuzzing}.
Fuzzing is the process of creating and running tests automatically with the intention of identifying  flaws~\cite{sutton2007fuzzing}.
Finding inputs that cause inappropriate program execution as a result of improper input data processing is the major goal of fuzzing~\cite{zhang2017s2f}.
Many fuzzing tools~\cite{restlerICSE2019,arcuri2019restful,\paperfifth,\paperfiftyfifth} rely on OpenAPI specification as it outlines how to use a REST API, including the types of requests it can handle, the possible responses, and the format of the possible responses.

%%%%%%%%%%%%%%%%%%%%%%%%%%%%%%%%%%%%%%%%%%%%%%%%%%%%%%%%%%%%%%%%%%%%%%%%%%%
\section{Related Work}
\label{sec:related}

In software engineering research, there are several surveys that summarize, categorize and analyze the current state-of-the-art on different research topics. 
Examples include
search-based test case generation~\cite{ABHP09},
search based mutation testing~\cite{silva2017systematic},
software-testing education~\cite{garousi2020software},
test case selection and prioritization using machine learning~\cite{pan2022test},
software testing effort estimation~\cite{bluemke2021software}, 
software robustness assessment~\cite{laranjeiro2021systematic},
oracle problem in software testing~\cite{barr2015oracle},
flaky tests~\cite{parry2021survey}
and
software engineering for AI-based systems~\cite{martinez2022software}.

Regarding the testing of web services, some old surveys (from 2009 and 2013) analyzed the challenges of testing service-oriented architectures~\cite{canfora2009service,bozkurt2013testing}.
However, as they are old, they do not represent the current state-of-the-art, especially considering the blossoming of research activities from 2017 on (which will be discussed in more details in Section~\ref{sub:ayear}).

There is currently another short survey on the testing of RESTful APIs~\cite{ehsan2022restful}, published in the \emph{Applied Sciences} journal in 2022.
Such survey is based on 16 published articles, addressing the following research questions:

\begin{itemize}
	\item RQ-1: What are the main challenges in generating unit tests for RESTful APIs?
	\item RQ-2: What are the code coverage concerns when it comes to testing RESTful APIs?
	\item RQ-3: What solutions are currently available to meet testing and unit test generation challenges?
	\item RQ-4: What support do solutions provide for authentication-enabled RESTful APIs’ testing and unit test generation?
\end{itemize}

Our survey is much larger, covering \totalPapers articles instead of just 16 (and all these 16 are included in our analyses).
Furthermore, compared to those 4 RQs listed in~\cite{ehsan2022restful}, we answer many more research questions from various perspectives (e.g., testing metrics, testing kinds, existing fuzzers and challenges), in order to provide a better overview of the current state-of-the-art in this domain.

%%%%%%%%%%%%%%%%%%%%%%%%%%%%%%%%%%%%%%%%%%%%%%%%%%%%%%%%%%%%%%%%%%%%%%%%%%%%

\section{Research Method} 
\label{sec:method}

%----------------------------------------------------------
\subsection{Research Questions}
\label{subsec:rqs}

To investigate current research in addressing REST API testing, we conduct a systematic literature review (SLR) to answer the following research questions from four perspectives:

\begin{itemize}
	\item Publication status in testing RESTful APIs
	\begin{itemize}
		\item \rqAYear
		\item \rqAVenues
		\item \rqAContri
	\end{itemize}
	\item Existing approaches in supporting automated testing of RESTful APIs
	\begin{itemize}
		\item \rqBMetric
		\item \rqBTechni
		\item \rqBTestKind
		\item \rqBCaseStudy
	\end{itemize}
	\item Available Tools for Testing RESTful APIs
	\begin{itemize}
		\item \rqCOpenSou
		\item \rqCNonResTool
		\item \rqCFeatures
	\end{itemize}
	\item Addressed and Open Challenges
	\begin{itemize}
		\item \rqDAdresChal
		\item \rqDOpenChal
	\end{itemize}
\end{itemize}

%----------------------------------------------------------
\subsection{Database and Search Queries}
\label{subsec:query}

To find relevant papers, we took advantage of seven databases as listed in Table~\ref{tab:databases}. 
These online repositories of peer-reviewed articles were selected based on their popularity and extent of relevance to software engineering research. 
These sources contain well-known conferences and journals in the field.
Existing surveys in software engineering research (\eg \cite{pan2022test,bluemke2021software} widely referenced them, as they offer a variety of authoritative publication venues in the field.

\begin{table}[!th]
	\small
	\centering
	\caption{Selected databases, with search queries and number of found articles}
	%	\vspace{-1\baselineskip}
	\label{tab:databases}
	\resizebox{1\textwidth}{!}{
		\begin{tabular} { l p{0.7\textwidth} c}\\ 
	\toprule 
	Name &  Search Query & Found Papers\\ 
	
	\midrule 
	IEEE & ("Document Title":restful OR "Index Terms":restful OR "Document Title":rest OR "Index Terms":rest)
	AND ("Document Title":test* OR "Index Terms":test*) AND ("Full Text .AND. Metadata":"white box" 
	OR "Full Text .AND. Metadata":"black box" OR "Full Text .AND. Metadata":fuzz* 
	OR "Full Text .AND. Metadata":"unit test*" OR "Full Text .AND. Metadata":"integration testing"
	OR "Full Text .AND. Metadata":"system test*" OR "Full Text .AND. Metadata":"end to end test*")  & 85\\ 
	
	\midrule

	ACM & (Title:(restful) OR Keywords:(restful) OR Title:(rest) OR Keywords:(rest)) AND (Title:(test*) OR Keyword:(test*)) AND (Anywhere:(white box) OR Anywhere:(black box) OR Anywhere:(fuzz*) OR Anywhere:("unit test") OR Anywhere:("unit tests") OR Anywhere:("unit testing") OR Anywhere:("integration test") OR Anywhere:("integration tests") OR Anywhere:("integration testing") OR Anywhere:("system test") OR Anywhere:("system tests") OR Anywhere:("system testing") OR Anywhere:("end to end test") OR Anywhere:("end to end tests") OR Anywhere:("end to end testing")) & 18\\ 
		
	\midrule 
	
	ScienceDirect & Title, abstract, keywords: (restful OR "rest api" OR "rest apis" OR "rest service" OR "rest services" OR "rest web services") AND (test OR testing OR tests)  & 17\\ 
	& Subjects: Computer Science & \\
	
	\midrule 
	
	Wiley & (test*) AND (restful OR "REST API" OR "REST APIs" OR "REST service" OR "REST services" OR "REST web service" OR "REST web services")
	& 2\\ 
	
	& Subjects: Computer Science & \\

	\midrule 
	
	Web of Science & (TI=(restful) OR AK=(restful) OR TI=("rest api"*) OR AK=("rest api"*) OR TI=("rest service"*) OR AK=("rest service"*) OR TI=("rest web service"*) OR AK=("rest web service"*)) AND (TI=(test*) OR AK=(test*))
	& 12\\ 
	
	\midrule
	
	MIT Libraries & Abstract contains: (restful OR "rest api*" OR "rest service*" OR "rest web service*") AND (test*) AND Any field contains: "black box" OR "white box" OR "unit test*" OR "integration test*" OR "system test*" OR "end to end test*" OR fuzz*
	& 19\\ 
	
	\midrule 
	
	Springer & restful AND test AND api & 90\\ 
	& Articles within Computer Science and Software Engineering/Programming and Operating Systems & \\ 
	
	\bottomrule 
\end{tabular}

	}
\end{table}

In order to achieve the objectives of this work and answer our research questions, we put together our search terms according to the specific format of each database. 
The queries were formulated around the concept of applying testing on RESTful APIs. 
In some databases, such as \textit{IEEE} and \textit{ACM}, to have more relevant results, we excluded the papers which do not contain at least one of the commonly used terms in the literature in any part of them (\eg by using \textit{Anywhere} in ACM) including \textit{black box}, \textit{white box}, \textit{fuzzing}, \textit{fuzzer}, \textit{unit test}, \textit{system test}, \textit{end to end test} and \textit{integration test}. 
The numbers of papers found as the result of search queries are shown in Table~\ref{tab:databases}. The search was conducted on March 1, 2022. 

Each repository had its own limitations to conduct advanced searches.
For example, unlike IEEE and ACM, it did not seem to be possible to formulate an advanced search query in Springer. 
At the time of writing this survey, there was no feature to limit the results based on different sections of the paper (\eg Title or Abstract) in Springer. 
Therefore, the only option we had was to do a broad search query and limiting it by subject (\ie \textit{Computer Science} and then \textit{Software Engineering/Programming and Operating Systems}) and type of the study (\ie \textit{Article}). 
In \textit{ScienceDirect}, there was a limitation for boolean operators. 
It did not allow more than 8 boolean operators, so we were not able to include terms such as \textit{SBST}. Another impediment with this repository was that it did not support wildcards.

%----------------------------------------------------------
\subsection{Paper Selection Criteria}
\label{subsec:selection}

Table~\ref{tab:inclusionCriteria} introduces the inclusion criteria we used to select papers. 
In general, we chose papers that are written in English, and that are related to testing in the domain of REST APIs. 
However, to be included, an article did not have to be exclusively in the domain of REST. 
It could be generally about web services, cloud services, etc., which might encompass RESTful APIs as well.
We excluded theses (e.g., MSc and PhD).
We also excluded existing surveys, as those are rather discussed in our Related Work (Section~\ref{sec:related}).

\begin{table}[!th]
	\small
	\centering
	\caption{Inclusion and Exclusion Criteria}
	%	\vspace{-1\baselineskip}
	\label{tab:inclusionCriteria}
	%{p{0.2\textwidth}CGC}
\begin{tabular} { l l }\\ 
	\toprule 
	&  Criterion \\ 
	\midrule 
	1 & The paper is in English \\ 
	2 & The paper is related to testing \\ 
    3 & The paper is about REST. It does not have to be exclusively about REST. \\ 
  	4 & The paper is not a thesis. \\
  	5 & The paper is not a survey or a systematic literature review. \\ 
	\bottomrule 
\end{tabular} 
\end{table}

%----------------------------------------------------------
\subsection{Snowballing}
\label{subsec:snowballing}

The idea of conducting this phase was to employ a hybrid search strategy to reduce the chance of missing important relevant papers.
To do so, we conducted forward and backward snowballing on June 14th, 2022.
Forward and backward snowballing refer to checking the reference list and citations of a paper, respectively. 
To find citations of a paper, we took advantage of \textit{Google Scholar}\footnote{http://scholar.google.com/}. 
By studying and evaluating them based on the defined inclusion criteria, we finally gathered \totalPapers papers in total, including \searchPapers ones which were initially found by search and also \snowballingPapers new ones by conducting snowballing.

There are several reasons why the number of papers found in this step are considerably larger than that of found during the initial search. 
First,
some of the papers were published after March 1st, 2022, which is the date we conducted the initial search. This group includes \papersAfterMarch papers. 
Second,
\papersFromOtherDatabases papers
were only available in sources that were not among those which had been selected to conduct the initial search (see Table~\ref{tab:databases}). These sources include \otherDatabases.
The third reason could be that the approach might not be specific to REST, and the term REST might be not present in either the title, abstract or keywords.
However, during the snowballing phase, if we found papers where the proposed approach was evaluated on REST APIs, then we include them in this survey.
In addition, there might exist some bugs in the search services provided by the article databases.
For instance, by taking a look at the paper~\cite{\paperthirtyeighth}, we found out that this should had appeared in results of the initial search as it fits our search query, but surprisingly did not. 
Then, we reported such issues to the service (\ie ACM), and those have been confirmed and fixed by now.

%----------------------------------------------------------
\subsection{Data Extraction}
\label{subsec:extraction}

Table~\ref{tab:data-extraction} contains the types of extracted data for each research question. 
We designed a spreadsheet in Google Sheets to put together collected data from the selected papers.
This included name of the paper, a unique given ID and information related to answering research questions.
Before conducting the data extraction,
we prepared a list of possible categories and relevant keywords to search for (e.g., \emph{black-box} and \emph{white-box}), and refined such selection while investigating each paper.
However, by starting to study the papers, we found some other possible values. 
For example, regarding RQ5, we found out that not all the papers use either black-box or white-box techniques, and also they  can be both supported at the same time (e.g.,\cite{\papersixth}).

To conduct the data extraction, the papers were divided between two of the 
authors. 
Each of them extracted the data based on the types shown in Table~\ref{tab:data-extraction}. Then, the results were double-checked by the other author. 
In case of any discrepancy, the issue was discussed and settled by the third author. 

\begin{table}[!th]
	\small
	\centering
	\caption{Type of Data Extracted Per Research Question}
	%	\vspace{-1\baselineskip}
	\label{tab:data-extraction}
	%{p{0.2\textwidth}CGC}

%\begin{tabular}{|*{4}{l|}}
\begin{tabular}{ l l}
	\toprule 	
	RQs     & Type of Extracted Data \\
	\midrule
	
	RQ1           & The number of papers per year \\ %\cline{1-3}
	
	RQ2            & The number of papers per venue \\ %\cline{1-3}
	
	RQ3              & 1) Types of contribution of the papers, and \\  

	    & 2) The number of papers per contribution type  \\  %\cline{1-3}

	RQ4            & Existing metrics which have been applied for guiding/evaluating REST API testing\\ %\cline{1-3}	

 	RQ5            & Existing testing techniques for REST APIs \\ %\cline{1-3}

	RQ6           & Types of testing which REST API testing approaches support  \\ %\cline{1-3}
		
	RQ7               & 1) Types of artifacts which have been used for conducting evaluation of REST API testing, and \\  
		
	    & 2) Locations and availability of the artifacts  \\ %\cline{1-3}
	
	RQ8            & 1) REST API testing approaches which have the open-source prototype, and  \\ 
	            &  2) Replicability of experiments conducted with the prototypes  \\ %\cline{1-3}
	
	RQ9 &  Existing non-research tools which have been applied for REST API testing \\
	
	RQ10            & Features which have been supported by research prototypes of REST API testing\\ %\cline{1-3}
	
	RQ11            & Research questions and hypotheses which have been studied in REST APIs testing \\ %\cline{1-3}
		
	RQ12           & Open challenges which have been identified in testing REST APIs \\ %\cline{1-3}
	\bottomrule
\end{tabular}

\end{table}

%%%%%%%%%%%%%%%%%%%%%%%%%%%%%%%%%%%%%%%%%%%%%%%%%%%%%%%%%%%%%%%%%%%%%%%%%%%%
\section{Status of Publications in REST API Testing}
\label{sec:survey}

To study the trends of research in REST API testing, 
in this section we report our investigation on existing studies of REST API testing with academic publications.
These findings are categorized in terms of time (RQ1), venues (RQ2), and main contributions (RQ3).

%----------------------------------------------------------
\subsection{\rqAYear}
\label{sub:ayear}

\begin{figure}[!th]
	\centering
	\includegraphics[scale=0.6]{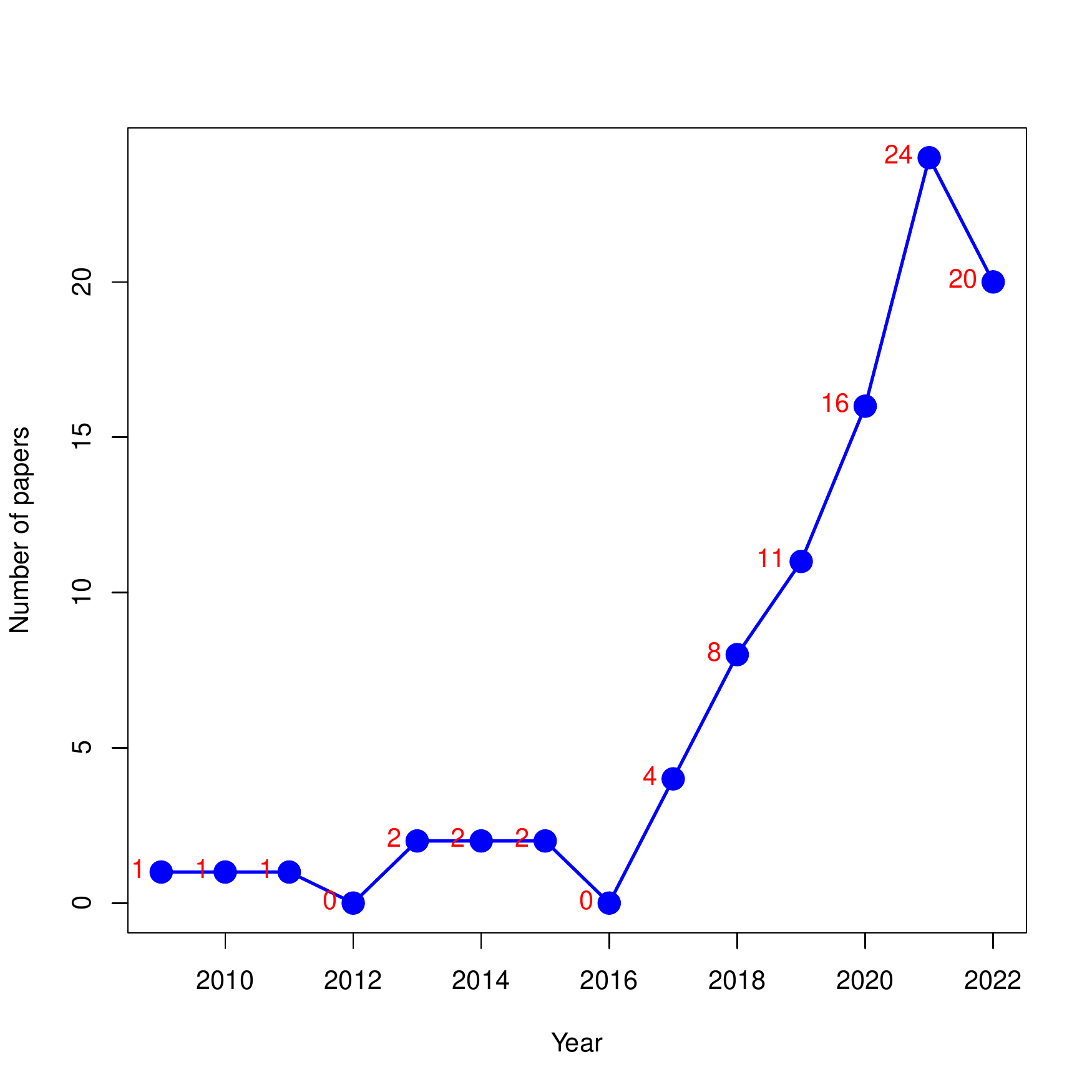}
	\vspace{-1.5\baselineskip}
	\caption{Number of selected papers (y-axis) per year (x-axis)}
	\label{fig:papers_per_year}
\end{figure}

Figure~\ref{fig:papers_per_year} illustrates the amount of papers published from 2009 to 2022. 
It shows that there has been an upward trend in the number of papers in the domain of RESTful APIs testing during recent years. 

There was not a considerable number of studies before 2017, as the number has fluctuated between 0 and 2. 
However, the quantity of published papers has increased dramatically from 2017 onward. 
In 2018, \paperstwentyeighteen papers were published and this number further increased and reached to the highest, \paperstwentytwentyone, in 2021. At the time of finalizing the selected papers for this survey (June 2022), only half of the year 2022 has passed. 
This is the reason why the number of papers published in 2022 is lower than that of 2021.

\begin{result}
	
	{\bf RQ1}: Since 2017, there has been a dramatic increase in the number of scientific studies on testing REST APIs. 
	
\end{result}

%----------------------------------------------------------
\subsection{\rqAVenues}

Papers selected for this survey have been published in a variety of venues. 
Table~\ref{tab:venues} gives information about the number of papers being published in each venue and its type. 
A venue can be of type conference, journal or other open-access repositories. 
The table is sorted in descending order, based on the number of papers.
To keep the table short, 
we represent a full name of a venue  only if:
(1) more than one paper is published in the venue 
\textit{or} 
(2) a journal article is published by one of the main publishers in the field of software engineering (i.e., ACM, Elsevier, IEEE, Springer, Wiley).
For others, we categorized them into \textit{Other Conferences} and \textit{Other Journals} based on the venue type.

Based on Table~\ref{tab:venues},~\totalPapers studies of REST API testing have been published in~\numberOfVenues various venues (\ie ~\numberOfConferences conference venues,~\numberOfJournals journals venues and~\numberOfOpenAccessRepos open access repository)  which have covered well-known top SE venues, such as TOSEM, TSE, EMSE and ICSE.
Regarding conference publications, there are in total \conferencePapers papers.
On top, with \otherConferences papers, there is \textit{Other Conferences}, which aggregates the amount of papers from conferences with a single publication.
The most common venues are
ICSE (\emph{ACM/IEEE International Conference of Software Engineering})
and
ICST (\emph{IEEE International Conference on Software Testing, Verification and Validation}),
with~\topVenueNumber and 6 papers.
In addition, \journalPapers articles have been published in journals, and the most frequent journal is TOSEM.
It is a peer-reviewed journal published by ACM, which stands for \emph{ACM Transactions on Software Engineering and Methodology}.
Besides, there also exist \openAccessRepoPapers papers which have been submitted to a non-peer reviewed open access repository,~\ie arXiv.

%\amid{Just be aware that the description of the venue is hardcoded.}

\begin{table}[!ht]
	\small
	\centering
	\caption{Number of papers per venue and their type}
	%	\vspace{-1\baselineskip}
	\label{tab:venues}
	\begin{tabular} { l l c r}\\ 
\toprule 
 & Venue &  Number of Papers & Venue Type\\ 
\midrule  
1 & \emph{Other Conferences} & \textbf{43} & Conference\\ 
2 & \emph{Other Journals} & \textbf{8} & Journal\\ 
3 & \emph{ICSE} & \textbf{7} & Conference\\ 
4 & \emph{arXiv} & \textbf{6} & Open-access Repository\\ 
5 & \emph{ICST} & \textbf{6} & Conference\\ 
6 & \emph{TOSEM} & \textbf{5} & Journal\\ 
7 & \emph{QRS} & \textbf{4} & Conference\\ 
8 & \emph{IEEE TSE} & \textbf{3} & Journal\\ 
9 & \emph{ESEC/FSE} & \textbf{2} & Conference\\ 
10 & \emph{GECCO} & \textbf{2} & Conference\\ 
11 & \emph{ISSTA} & \textbf{2} & Conference\\ 
12 & \emph{Applied Soft Computing} & \textbf{1} & Journal\\ 
13 & \emph{EMSE} & \textbf{1} & Journal\\ 
14 & \emph{IEEE Software} & \textbf{1} & Journal\\ 
15 & \emph{IEEE TSC} & \textbf{1} & Journal\\ 
\bottomrule 
\end{tabular} 

	%	\resizebox{!}{0.5\textheight}{
		%\input{tables/rq2}}
\end{table}

\begin{result}
	{\bf RQ2}: From~\firstYear to~\lastYear, scientific work on testing RESTful APIs was published in~\numberOfVenues different venues, where ICSE/ICST and~\topJournal were the most frequent conferences and journal, respectively.
\end{result}

\subsection{\rqAContri}
\label{subsec:contributionType}

To better study what contributions have been made in existing studies, we
defined \NumberOfContributionTypes different categories to classify main contributions of selected papers, as follows.

\begin{itemize}
	\item \textbf{\NA}
	
	This category includes papers which propose a new tool, algorithm, framework or method for automated testing of REST APIs and its integrated extensions for enhancing the automated approach. 
	For example, regarding automated test case generation for REST APIs, 
	Arcuri proposed the Many Independent Objective (MIO) algorithm specific to white-box system level test generation~\cite{arcuri2018test}, and enabled search-based software testing (SBST) of REST APIs with  MIO~\cite{arcuri2019restful},  implemented as an open-source fuzzer, named \evo.
	To further improve the SBST of REST APIs, SQL handling~\cite{\paperthirtyninth} and testability transformation~\cite{\paperfiftieth} were integrated into \evo.
	However, such extension techniques can be used outside of REST APIs, i.e., applied for testing of other domains.
	Another example for test data generation is
	\textit{ARTE}, which is an approach presented in~\cite{\paperfortyfirst} to automatically extract realistic data inputs for REST APIs from knowledge bases (\eg DBpedia) by taking advantage of a number of techniques, including natural language processing, search-based, and knowledge extraction.
	To make ARTE fully automated, it is integrated into RESTest~\cite{\papereleventh}.
	%Zhang \etal have proposed a new method in \cite{\papertwentyfifth} for automated system test generation which is also implemented as an extension to \evo~\cite{\papertwentythird} that is a tool for automatically generating system tests for REST APIs. 
	%LT-MOSA~\cite{\papertenth} is an example of a new algorithm which is aimed at generating system-level test cases.

	\item \textbf{\NP}
	
	Studies which, instead of proposing a new
	approach for directly achieving automated testing,
	help to %automate an existing approach for testing REST APIs, or contribute toward enhancing an already automated approach, 
	analyze testing related aspects (such as coverage metrics) and identify potential solutions for testing of REST APIs,
	fall into this category. 
	For instance, Marculescu~\etal~\cite{\paperfiftyfirst} studied faults in RESTful APIs selected from EMB repository~\cite{EMB}, and proposed a taxonomy of the faults identified in the REST API with \evo~\cite{\paperfiftyfirst}.
	Martin-Lopez~\etal~\cite{\papertwentyfourth} defined a set of coverage metrics based on API schema in the context of black-box testing of REST APIs.
	Katt and Prasher~\cite{\papertwentysecond} identified potential security threats in REST APIs and proposed corresponding quantitative security assurance metrics.
	In addition, Soni~\etal~\cite{\paperseventysixth} proposed a framework for allowing mocking external dependencies of Java REST APIs.
	However, this mocking solution is for unit testing, and it could be potentially adopted to automated testing approaches  for enabling additional handling of external services.

	\item \textbf{\EA}
	
	This group of papers focus on comparing existing approaches by conducting empirical evaluations.
	For instance, the study conducted by Corradiani \etal in ~\cite{\paperninth} automated black box approaches for testing REST APIs have been compared by conducting an empirical comparison on them.
	There also exist studies for comparing existing fuzzers for REST APIs~\cite{\paperfortyfourth} and analyzing open problems~\cite{\paperfortieth} with the fuzzers.
	
	% Move it to NP
	%\item \textbf{\EI}
	%This type of contribution encompasses the papers in which testing methods have been applied to evaluate a group of REST APIs. The only paper falling in this category applies search-based software testing methods on different RESTful API case studies selected from EMB repository~\cite{EMB} to present a taxonomy of the faults identified in REST API with \evo~\cite{\paperfiftyfirst}.
	
	\item \textbf{\TD}
	
	Papers which mainly focus on implementation of a testing tool, or show how a tool can be used, fall into this category.
	For instance, \restats is introduced in \cite{\paperthirtythird}, which is a tool to compute coverage metrics for black-box testing of REST APIs.
	This tool adopts the coverage metrics proposed by Martin-Lopez \etal~\cite{\papertwentyfourth}.
	
	%	(\eg ~\cite{\paperthirtythird})
	\item \textbf{\PI}
	
	These papers do not include any conducted study.
	Instead, they suggest a new idea or plan to carry out research on the testing of REST APIs.
	For instance, Martin-Lopez in ~\cite{\paperthirteenth} was planning to develop a framework for specification-driven testing that will automatically create complex test cases for Web APIs and also intelligent programs (called “bots”) which can generate a large number of inputs.
	
	%	(\eg ~\cite{\paperthirteenth})
\end{itemize}

\begin{figure}[!th]
	\centering
	\includegraphics[scale=0.6]{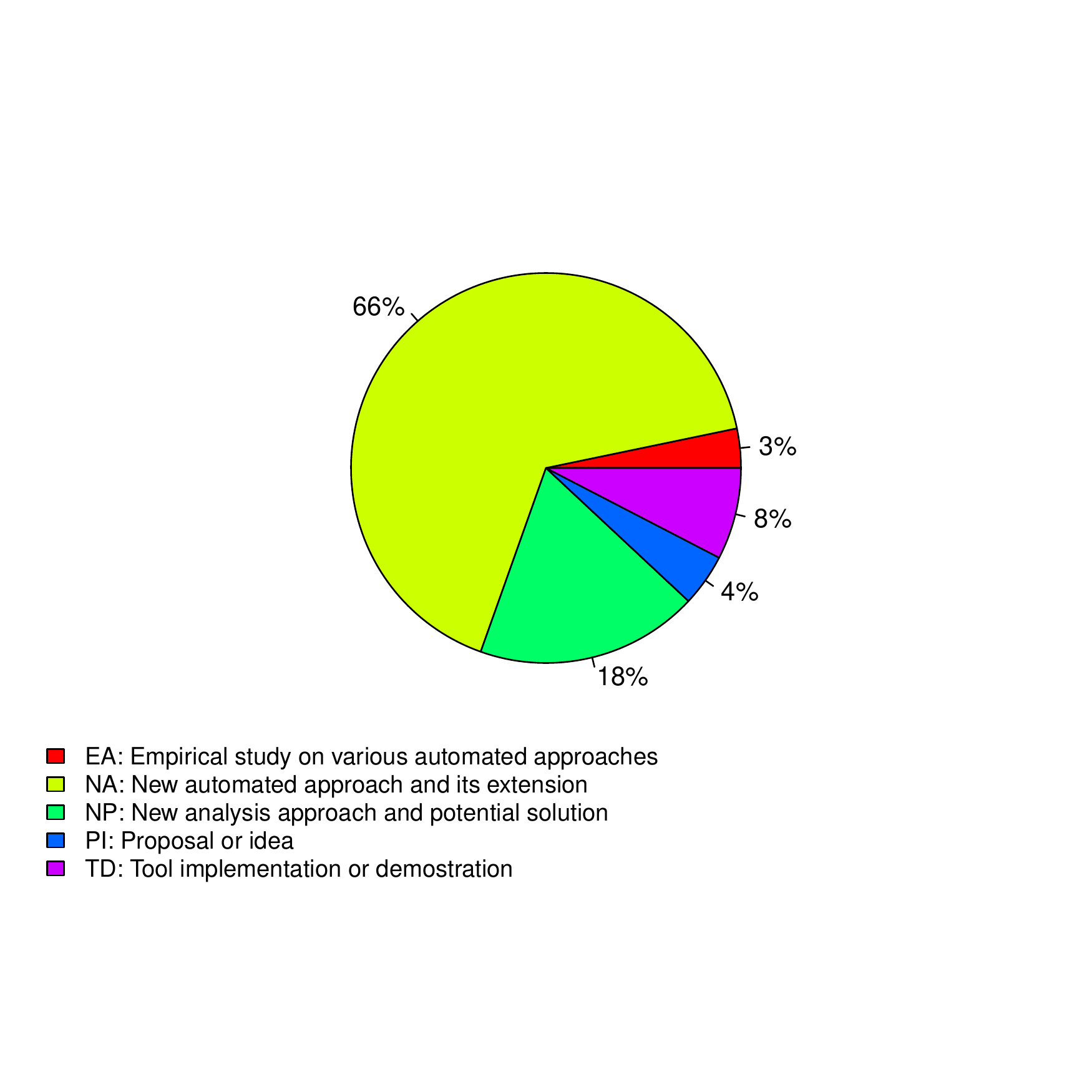}
	\vspace{-4\baselineskip}
	\caption{Types of paper contributions and their frequencies}
	\label{fig:rq3-pie}
\end{figure}

The data provided in the form of pie-chart in Figure~\ref{fig:rq3-pie} shows the share of selected papers based on their main contribution.
%It can be seen that more than two-third of the papers are proposing a new automated approach.
``\NA'' has the highest percentage with \NAPercent.
``\NP'' is the second largest category and comprises of almost one-fifth of the papers by \NPPercent.
The three other categories are much smaller (\eg~ \EAPercent for ``\EA'').

More details on the different categories and type contributions are presented and analyzed in the following
research questions.

%\andreaAddressed{give more info, or provide sentence stating this will be anaylzed/discussed in more details later in the paper}

\begin{result}
	{\bf RQ3}: \NumberOfContributionTypes different contributions were found from the main contributions of the selected \totalPapers papers.
	%Almost two-thirds 
	\NAPercent of the papers propose a \NAlower for testing RESTful APIs, while \NPPercent of the papers are \NPlower of REST API testing.
\end{result}

%%%%%%%%%%%%%%%%%%%%%%%%%%%%%%%%%%%%%%%%%%%%%%%%%%%%%%%%%%%%%%%%%%%%%%%%%%%%
\section{Approaches to Testing REST API}
\label{sec:approaches}

One of the main objectives of this survey is to identify existing testing approaches which have been developed to be automated in the context of REST APIs.

\begin{figure}
	\centering
	\includegraphics[width=0.7\linewidth]{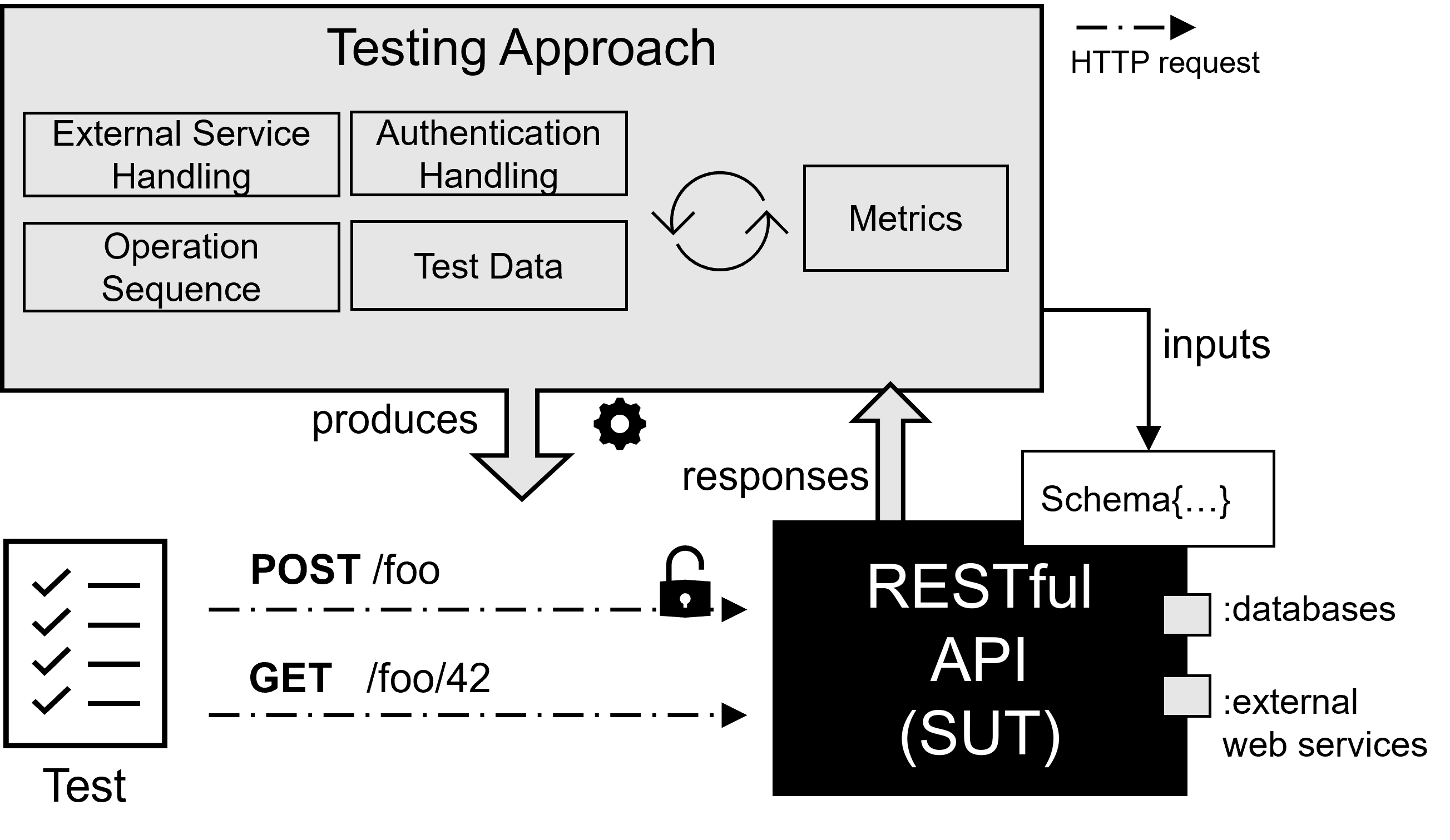}
	\caption{High-Level View of REST API Testing}
	\label{fig:approaches}
\end{figure}

Figure~\ref{fig:approaches} represents a high-level abstraction of testing
of REST APIs.
A RESTful API typically exposes a schema about how to access the web service.
The schema could be represented in various ways (e.g., JSON, XML, formal model), and 
one popular technique to define the schema is OpenAPI, as discussed in Section~\ref{subsec:openapi}.
As the schema defines the structure of the resources handled by SUT and the available actions to access these resources~\cite{Swagger,zhang2021resource}, it is often used as an input to test the REST API, e.g., define metrics as test criteria, automatically generated tests (referred as the term \textit{fuzzing}~\cite{godefroid2020fuzzing}).
The test generation is typically guided by heuristics aimed at optimizing those metrics.
Such metrics could be linked with behaviors of the SUT and data produced by the SUT offline or at runtime.
A test for a REST API can be regarded as a sequence of HTTP requests, e.g., a sequence of \texttt{POST /foo} and \texttt{GET /foo/42} as shown in Figure~\ref{fig:approaches}.
In order to perform actions on the SUT, each request has concrete values (referred as test data, e.g., 42) for its parameters, such as \textit{path parameters}, \textit{query parameters}  and \textit{body payload}, if specified in the schema.
In addition, endpoints of REST APIs might be restricted with \textit{authentications}, and the REST API could also connect to \textit{databases} and \textit{external web services} (as shown in Figure~\ref{fig:approaches}).
How to configure the authentication and handle such external services are also part of REST API testing.

To study the existing approaches of REST API testing, we designed RQs 4--7 and reported results of our investigation on metrics which include test criteria and heuristics (Section~\ref{subsec:effectinveness-metrics}),
techniques (Section~\ref{subsec:technique}),
kinds of testing
which have be applied for (Section~\ref{subsec:testingkind}),
and available artifacts used in the empirical evaluations (Section~\ref{subsec:artifacts}).

%----------------------------------------------------------
\subsection{\rqBMetric}
\label{subsec:effectinveness-metrics}

\begin{figure}
	\centering
	\includegraphics[width=0.7\linewidth]{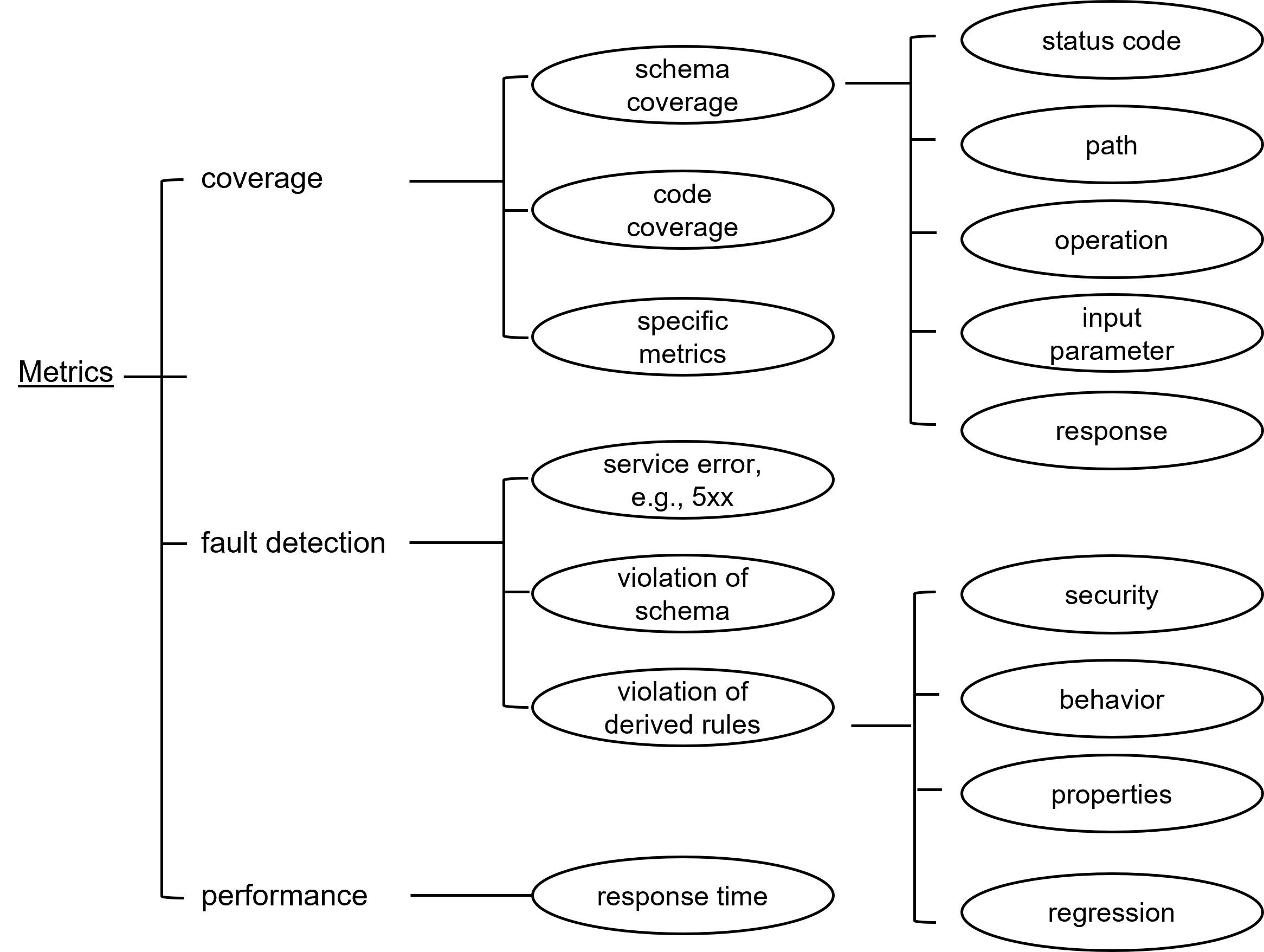}
	\caption{Metrics employed for REST API Testing in the literature}
	\label{fig:metrics}
\end{figure}

\begin{figure}
	\centering
	\includegraphics[width=1\linewidth]{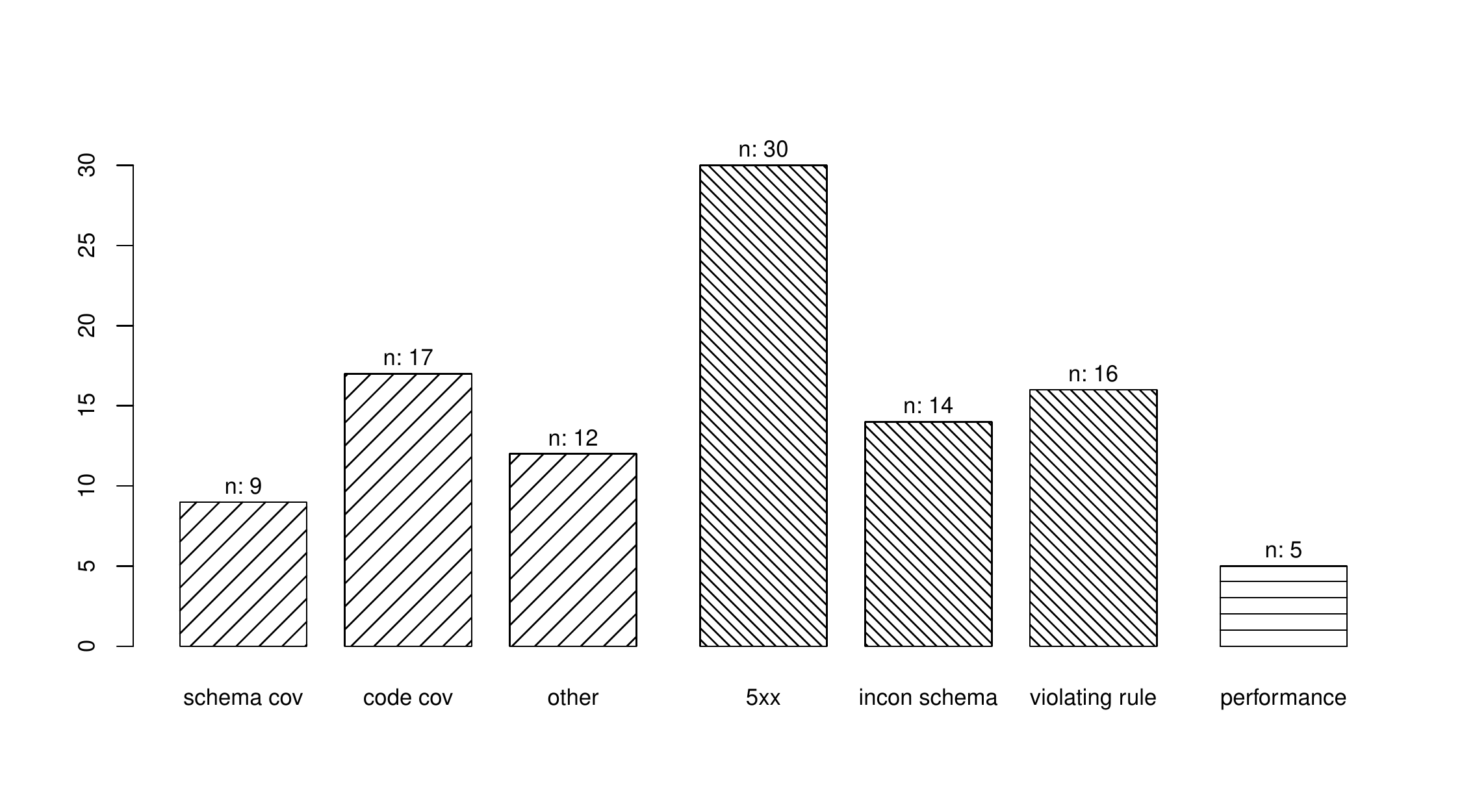}
	\caption{Statistics of metrics employed in automated testing of REST API}
	\label{fig:metrics_results}
\end{figure}

With this SLR, based on the selected papers, we classified metrics into three types, i.e., %\textit{success/failure},
\textit{coverage} criteria, \textit{fault detection} and \textit{performance}, as shown in Figure~\ref{fig:metrics}, and statistics of each type are shown in Figure~\ref{fig:metrics_results}.
Those are discussed next in more details.

%% might remove, since it does not provide new insights

%----------------------------------------------------------
%\subsubsection{Success/Failure}
%
%REST is built on the top of the HTTP.
%Based on HTTP responses, success or failure of HTTP requests could be identified with the status code.
%Such success and failure results are one of metrics applied in fuzzing REST APIs, such as \textit{APIs with status code 2xx} in~\cite{\paperfirst,\paperseventeenth}.
%
%\man{might remove this category, check at the end}

%----------------------------------------------------------
\subsubsection{Coverage Criteria}

%----------------------------------------------------------
\paragraphfourth{(1) schema related coverage}

API schemas are main inputs to guide how to access the SUT and define what responses should be returned.
For instance, as an example of the schema defined with the OpenAPI (Figure~\ref{fig:openapi}), an endpoint could be defined under a \texttt{URI path} with an HTTP verb (e.g., \texttt{POST}, \texttt{GET}), input parameters (e.g., \texttt{Path Parameter}) and possible responses (e.g., status code, response body).
Then, there exist several black-box coverage metrics which are defined based on those elements~\mCovSchema:
\begin{itemize}
	\item \textit{HTTP status code} could reflect a result of processing the request in the SUT, such as 2xx often represents a successful request.
	A set of testing criteria has been defined in order to compute a coverage of status codes which are returned during testing, for each endpoint~\cite{\paperfirst,\papersixth,arcuri2019restful,\papertwentyfirst,\papereleventh,\paperseventeenth}.
	\item \textit{path} provides info to access the API, i.e., the full URI to access an endpoint could be constructed by \textit{base path} plus \textit{path}.
	For instance, Martin-Lopez~\etal~\cite{\papertwentyfourth} defined \textit{path coverage} which assesses a number of paths accessed by the generated tests out of the total available paths in the schema.
	Banias~\etal~\cite{\papersecond} evaluated \textit{paths tested} by considering success and failure responses received by requests to the path.
	Ed-douibi~\etal~\cite{\papertwentysixth} reported \textit{endpoint coverage}, and an endpoint is considered as covered only if all of its operations are covered.
	\item \textit{operations} are exposed to make requests (i.e., HTTP verb with path) for performing actions on the services.
	%\andrea{bit confused. what is difference between `path' and `operations'?}
	As discussed, a test is regarded as a sequence of the operations.
	To evaluate REST API testing approaches, Banias~\etal reported a number of operation tested in~\cite{\papersecond}.
	%one of operation related coverage is to measure a percentage of operations which have received successful responses for a API~\cite{arcuri2019restful}.
	%
	\item \textit{input parameter} is info required to set when making a request. 
	The parameter could be different with various types and constraints (e.g., \textit{required}, \textit{minimum}). 
	Then the input parameter based metrics are defined to assess if various values for the parameters have been examined during testing (e.g., each boolean parameter should had been evaluated with both values \texttt{true} and \texttt{false}).
	The generation could also be guided by the metrics, e.g., Banias~\etal~\cite{\papersecond} defined various configurations to generate tests with a consideration of the \textit{required} property of the parameters.  
	\item \textit{response} defines a list of possible responses to return per operation. 
	Metrics relating to responses are used to examine whether various responses have been obtained (e.g., for an enumeration element in a returned body payload, coverage metrics would check if every single item in the enumeration has been returned at least once).
	Response body property coverage is reported to assess API schema based REST API testing approach in~\cite{\papersecond}.
\end{itemize}

In addition, Martin-Lopez~\etal~\cite{\papertwentyfourth} proposed 10 coverage metrics based on the schema that have been applied for assessing REST API testing approaches in~\cite{\papersecond}.
The 10 coverage metrics enable assessments of generated tests in fuzzing REST APIs with different inputs and outputs, such as \textit{parameter coverage, content-type coverage, response body properties coverage}.
The metrics were also enabled in fuzzers, such as \HsuanFuzz~\cite{\paperfiftysecond} and \restest~\cite{\papereleventh}.
Restats~\cite{\paperthirtythird} is a test coverage tool for assessing given tests based on the coverage metrics. 
In an empirical study~\cite{\paperninth}, such schema related coverage metrics were employed to compare black-box fuzzers of REST APIs.

%----------------------------------------------------------
\paragraphfourth{(2) code coverage}

Code coverage is one typical white-box criterion 
%\andrea{criterion is singular, criteria is plural} 
for evaluating testing approaches.
It was applied as well to evaluate REST API fuzzers for both black-box~\cite{\papersixtyfourth,\papertwentyfirst} and  white-box~\cite{\paperthird,\papersixth,\paperthirtyeighth,\paperthirtyninth,\paperfortieth,\paperfortyfifth}.  %\man{complete the list of the papers}.
%Regarding REST API testing,  the web services could be deployed remotely when performing black-box testing.
%However, applying code coverage is required to obtain the source code as a prerequisite.
Among different code coverage criteria, line/statement coverage is one of the most common and supported ones in industry-strength coverage tools (e.g., JaCoCo for Java programs).

Code coverage was also used as an evaluation metric to assess the black-box REST API testing approaches~\cite{\paperfortyeighth,\papersixtythird,\papertwentyfirst,\paperfiftysecond}.
For instance, in~\cite{\papertwentyfirst}, Atlidakis~\etal reported accumulated code coverage achieved along with a number of requests to be executed.
To collect the code coverage, \texttt{TracePoint} hook was configured in Ruby classes to trace the execution in this study.
However, black-box testing of REST API could be performed on remote public APIs which are closed source software.  
The code coverage in this context is applied only for evaluation purposes.

Code coverage could also be employed as a criteria to automate fuzzing, which is considered as white-box testing.
\evo enables runtime code coverage (with heuristics to maximize it) automatically during fuzzing with code instrumentation.  
However, such automated collection of runtime code coverage is specific to the programming language, and  \evo has enabled it for JVM (e.g., Java and Kotlin)~\cite{arcuri2019restful} and JavaScript~\cite{js2022}.
In addition, \Pythia~\cite{\papersixtyfourth} employs the metrics of code coverage to produce tests. 
But, to enable such code coverage collection, it needs a pre-manual step to extract code info (such as block location) of the SUT.

% Restler

%----------------------------------------------------------
\paragraphfourth{(3) other metrics}

Besides schema-based and code coverage, there also exist other metrics specific to proposed approaches~\mCovSpec.

%\man{add a table to list the specific metrics if there are not so many at the end}

For instance, in \RESTler~\cite{\papertwentyfirst}, a \textit{grammar} is derived based on the API schema for driving following test sequence generation, e.g., selecting next HTTP request based on the derived producer-consumer dependencies among the endpoints.
A specific metric employed in this approach is based on the \textit{grammar}, i.e., \textit{grammar coverage}.
Martin-Lopez~\etal~\cite{\paperfiftysixth,\paperninetyseventh,\paperninetysecond} defined \textit{inter-parameter dependencies} which represent constraints referring to two ore more input parameters. 
Such dependencies are also used in \RestCT to generate test data~\cite{\paperfortythird}.
Alonso \etal~\cite{\paperfortyfirst} enabled test data generation with realistic inputs using natural language processing and knowledge extraction techniques.
The performance of the data generation was evaluated from a percentage of \textit{valid} API calls (i.e., 2xx status code) and \textit{valid} inputs (i.e., \textit{Syntactically valid} and \textit{Semantically valid}).
%\RestTest~\cite{\papereighth} defines \textit{inter-parameter dependencies}
In~\cite{\papereighteenth}, Godefroid~\etal introduced the \textit{error type} metric which is a pair of error code and error message.
The metric was used to guide data fuzzing of REST API~\cite{\papereighteenth}, i.e., maximizing \textit{error type coverage}.

Chakrabarti and Rodriquez~\cite{\paperthirtyfirst} defined POST Class Graph (PCG) to represent resources and connected relationships among them, then tests could be produced by maximizing coverage of the graph.
UML state machine was applied in a model-based testing of REST APIs~\cite{\paperfortysixth}.
Metrics specific to model such as state coverage and transition coverage are used to guide test generation.

Lin~\etal~\cite{\paperfortyeighth} constructed tree-based graph to analyze resource and resource dependency based on the API schema and responses. 
Then test cases could be generated based on the graph with tree traversal algorithms.

To analyze security issues existing in REST API, Cheh and Chen~\cite{\papersixtyfirst}  analyzed levels of \textit{sensitivity} of data fields and API calls, and also defined \textit{exposure level} that calculates a degree of such data fields and API calls exposed to potential attacks. 
%\evo~\cite{\paperthird} employs derived \textit{resource dependency} to produce a sequences of requests in order to perform actions on manipulating resources.
%\andrea{is this really for specific metrics?}

%Richardson Maturity Model (RMM)

%----------------------------------------------------------
\subsubsection{Fault Detection}

%----------------------------------------------------------
\paragraphfourth{(1) service error}

The status code 5xx has been applied to identify potential faults in REST API testing~\mFaultFivexx.
With HTTP, the status code 5xx indicates errors caused by the server, and the request could not be processed until the server has been fixed.
For instance, 500 status code is generic to represent that an internal server error occurs when performing the given request.
%However, 5xx status code should be avoided and fixed before deploying the service in production.
%To detect faults in REST APIs, a number of 5xx status code is typically employed in its testing.
%\andrea{what stated here wasn't fully correct. i commented it out}
The status 503 is more specific, stating that the service is unavailable, e.g., due to down for maintenance or the server is overloaded.

In most HTTP frameworks, when there is a crash in the business logic due to some faults (e.g., an unhandled null-pointer exception), the entire server does not crash.
In these cases, the server would still reply to the incoming HTTP request, responding with a status code of 500.
Therefore, 500 status codes in the responses can be used as an oracle to detect faults in RESTful APIs~\cite{\paperfiftyfifth}. %\andrea{cite Bogdan TOSEM paper here}.
However, not all 500 status codes are related to software faults.
For example, if the API is connecting to a database, and the database is currently down, the server would not be able to complete the request.
In such a case, returning a 500 status code would be correct, although no software fault in the API is involved.

%----------------------------------------------------------
\paragraphfourth{(2) violation of schema}

The API schema (such as OpenAPI) defines response syntax for each operation~\mFaultSchema, e.g., status code and response body.
Actual responses should be always consistent with the syntax specified in the schema.
Thus, any inconsistency between the actual response and the syntax could be regarded as faults in the REST API.
For instance, 
Viglianisi~\etal defined such oracles in \RestTestGen~\cite{\paperseventeenth} by using 
an OpenAPI library\footnote{\url{https://github.com/bjansen/swagger-schema-validator}} to identify the mismatched responses.
In QuickRest~\cite{\papersixteenth}, Karlsson~\etal formulated such consistency as properties. 
\evo also reports the number of such faults at the end of fuzzing~\cite{\paperfiftieth}.

%----------------------------------------------------------
\paragraphfourth{(3) violation of defined rules}

Service errors (based on status code) and violation of schema (based on OpenAPI) are general oracles for fault finding in the context of REST API.
Besides, there also exist some oracles to identify faults based on rules which characterize the REST APIs in terms of \textit{security}, \textit{behavior}, \textit{properties} and \textit{regression}~\mFaultRule (see Figure~\ref{fig:metrics}).

\paragraphfifith{\textbf{Security}}
As web services, security is critical for REST APIs. 
To enable test oracle relating to security,  
Atlidakis~\etal~\cite{\paperfourteenth} proposed a set of rules which formalize desirable security related  properties of the REST APIs.
Any violation of the rules is identified as potential security related bugs in the SUT.
The rules are mainly defined based on assessing accessibility of resources, such as \textit{use-after-free rule: if a resource has been deleted, it must not be accessible anymore}.

% check OWASP Application Security Verification Standard
Katt and Prasher~\cite{\paperthirtyeighth} proposed an quantitative approach to measure the kinds of security related metrics (i.e., vulnerability, security requirement and security assurance) for web services.
To calculate the metrics, test cases are defined for validating whether the SUT meets security requirements and any kind of vulnerabilities exists.
%
%penetration testing
Masood and Java~\cite{\paperthirtyseventh} identified various kinds of vulnerabilities which could exist in REST APIs, such as JSON Hijacking.
Such vulnerabilities could be detected with both static analysis and dynamic analysis techniques. 
Barabanov~\etal~\cite{\paperfortyninth} proposed an approach specific to detection of Insecure Direct Object Reference (IDOR)/Broken Object Level Authorization (BOLA) vulnerabilities for REST APIs.
The approach analyzes the OpenAPI specification by identifying its elements (such as parameters) relating to IDOR/BOLA vulnerabilities, then generates tests for verifying the API with such elements using defined security rules.
%
%
%Mai~\etal~\cite{\paperninetysixth} defined metamorphic relations with security properties to detect potential vulnerabilities existing in the web system.
%
%
Zha~\etal~\cite{\papersixtyeighth} collected \textit{Common Sense Security Policies} (CSSP), such as \textit{Access control}, \textit{URL spoofing} and private messages for Team Chat system and defined CSSP violation scenarios.
Security and privacy risks can be identified if an API under the CSSP violation scenarios can still work, e.g., return a valid response.
Barlas~\etal~\cite{\papersixtysixth} studied regex-based denial of service (ReDoS) vulnerabilities lead by handling of input sanitization in web services.
The vulnerabilities is allowed to be identified by verifying consistency between client-side and server.

\paragraphfifith{\textbf{Behavior}}
Based on the provided API schema,
Ed-douibi~\etal~\cite{\papertwentysixth} defined two rules to generate \textit{nominal} test cases and \textit{faulty} test cases.
The \textit{nominal} test cases take the inputs inferred based on examples or constraints in the schema, and the successful response is expected to return (e.g., assert that the status code is 2xx). 
Regarding the \textit{faulty} test cases, it takes invalid inputs (e.g., missing required parameters, a string for a number parameter, a string violating its defined pattern), and the client error response is expected to return (e.g., assert that the status code is 4xx). 

Liu~\etal~\cite{\paperseventyninth} constructed five constraints of REST guidelines with models.
Then such models could be used to verify design models of REST APIs.
Any violation of the constraint models is considered as a potential defect in its architecture design.

Pinheiro~\etal~\cite{\paperfortysixth} defined UML state machines for modeling behaviors of REST APIs. 
%Earle~\etal~\cite{\papersixtieth} derived the state machine
The actual behavior (by executing the tests) should be consistent with the model (e.g., guard condition, invariant) as expected. 
Any inconsistency is recognized as potential faults of the SUT.  

\paragraphfifith{\textbf{Properties}}
Most of the HTTP methods are idempotent, i.e., GET, PUT, DELETE, HEAD, OPTIONS and TRACE.
For such methods, a result of executing the method is independent with the number of repeated times, i.e.,
executing the method multiple times would not change the result as executing it one time.
Thus, assertions could be defined to check the idempotency~\cite{\papertwelfth}, e.g., executing multiple identical GET should result in the same response, after a successful DELETE, responses of all following identical DELETE should be same.
Connectedness is examined in~\cite{\paperthirtyfirst} that refers to accessibility among resources.
For instance, assume that resource X owns resource Y, when perform a GET collection on Y referring to X, all available Y should appear in the response, otherwise the REST API is not ``connected''.    

The REST API could apply HATEOAS (Hypermedia as the Engine of Application State).
Then the response might contain hypermedia links for accessing itself or other resources.
For such responses, Vassiliou-Gioles~\cite{\papertwelfth} defined assertions for validating availability of links in its response.
In~\cite{\paperninetyfourth,\papereightieth}, Vu~\etal also proposed a model-based approach which enables formalization of hypermedia behaviors of the REST API with $\varepsilon$-NFA, and the model could allow an identification of the faults by checking whether the SUT complies with it~\cite{\papertwentieth}.
Fertig and Braun~\cite{\papertwentyninth} also verified the REST APIs based on hypermedia constraints using model-based approaches.

\textit{Metamorphic relation}s capture necessary properties which the SUT should hold with multiple executions.
To enable metamorphic testing of REST APIs, there exist several works to identify such \textit{metamorphic relation}s of the web services with abstract Metamorphic Relation Output Patterns (MROPs) ~\cite{\papertwentyseventh,\paperseventyeighth} (such as equivalence, disjoint) or specific properties of the API~\cite{\paperninetieth}.
Then, faults could be detected by checking whether responses among the multiple requests conform to the identified relations.   

\paragraphfifith{\textbf{Regression}}
Gazzola~\etal~\cite{\papereightyeighth} enables monitoring and tracing of the microservices in order to record their execution.
Such recorded execution slices can be abstracted and considered as a metric for generating regression tests, i.e., verify whether the same response could be received with the same request in the further version of the SUT.
Godefroid~\etal~\cite{\paperthirtysixth} employed \Restler to produce tests, then enabled detection of regression faults by comparing behaviors with the same tests among different versions of the REST APIs.
%\man{TODO add differential paper}

%----------------------------------------------------------
\subsubsection{Performance Metrics}

Performance related metrics are also important for REST API testing, as the SUT provides  services over the network.
In~\cite{\papersecond}, Banias~\etal measured average response time of the requests generated by different strategies.
Fertig and Braun~\cite{\papertwentyninth} discussed potential solutions to enable performance testing with model-based approaches and existing techniques (such as Apache JMeter\footnote{https://jmeter.apache.org/}).
\Schemathesis~\cite{\paperfiftyfifth} defined the proprieties relating to performance metrics, i.e., identify slow response and request amplification with configured thresholds.
Bucaille~\etal~\cite{\paperfortysecond} developed a testing framework which can assess and monitor performance-related properties, such as latency, by sending requests from various geographical locations using different cloud service providers.
%However, with this SLR, there does not exist an approach which facilitates automated testing of REST APIs with considerations of performance metrics.

\begin{result}
	{\bf RQ4}: Fault detection was the most applied metric in REST API testing that can be identified based on 5xx status codes, the given API schema and defined rules.
	Coverage criteria were the second widely applied metrics that measure the degree to which aspects of the REST APIs are tested.
	In the context of REST APIs, besides the traditional code coverage in  white-box testing, schema based coverage was mainly employ in the black-box testing.
	Performance metrics were rarely investigated in  REST API testing.
\end{result}

%----------------------------------------------------------
\subsection{\rqBTechni}
\label{subsec:technique}

To answer this question, we conducted further analysis on \NANumber (out of \totalPapers) papers whose contributions are categorized as ``\NAlower'' (see Section~\ref{subsec:contributionType}).

%----------------------------------------------------------
%\subsubsection{Offline vs Online Testing}

%----------------------------------------------------------
%\subsubsection{Test Data Generation}
%
%In the context of REST API testing, fuzzing of REST APIs requires to generate inputs for operations in order to make requests to SUT (see Figure~\ref{fig:approaches}).
%What data are generated could directly affect effectiveness of the fuzzers.
%With the SLR, we found that three manners to produce the data, i.e., \textit{static (offline)}, \textit{dynamic (offline)} and \textit{dynamic (online)}, as shown in Figure~\ref{fig:testdata}.
%
%\begin{figure}
%	\centering
%	\includegraphics[width=0.7\linewidth]{test_data_generation.png}
%	\caption{Test data generation in Fuzzing REST APIs}
%	\label{fig:testdata}
%\end{figure}
%
%
%\paragraphfourth{(1) static offline}
%
%\paragraphfourth{(2) dynamic offline}
%
%
%\cite{\papereighth}
%
%\paragraphfourth{(3) dynamic online}
%
%
%In~\cite{\papernineteenth}, Zhao~\etal presents an approach which employs a parser (i.e., ANTLR\footnote{https://www.antlr.org/}) to derive semantics of input parameters and semantic relationships of the REST API based on the OpenAPI specification, then generates test data based on the derived semantic info.

%----------------------------------------------------------
%% Man: important papers summary

%----------------------------------------------------------
\subsubsection{Black-box and White-box Testing}

\begin{table}
	\small
	\centering
	\caption{Results of techniques for automated testing of REST APIs}
	\begin{tabular}{  l r l r p{8cm}}
\toprule 
 Black-/White- box & \# & Tehcniques & \# &Papers\\ 
\midrule 
\emph{Black Box}&45&search&1&\cite{ \papereightyfifth }\\ 
&&model&8&\cite{ \papereleventh,\papertwentieth,\papertwentysixth,\papertwentyninth,\paperthirtieth,\paperfortysixth,\paperseventyninth,\papereightieth }\\ 
&&property&9&\cite{ \papersixteenth,\papertwentyseventh,\paperthirtieth,\paperthirtyfirst,\paperfiftyfifth,\papersixtieth,\paperseventyeighth,\paperninetieth,\paperninetysixth }\\ 
&&others&28&\cite{ \paperfirst,\papersecond,\paperfifth,\papereighth,\papertwelfth,\paperfourteenth,\paperfifteenth,\paperseventeenth,\papereighteenth,\papernineteenth,\papertwentyfirst,\paperthirtysixth,\paperfortyfirst,\paperfortysecond,\paperfortythird,\paperfortyeighth,\paperfortyninth,\paperfiftysecond,\paperfiftyfourth,\papersixtyfirst,\papersixtysecond,\papersixtythird,\papersixtysixth,\papersixtyeighth,\papersixtyninth,\papereightysixth,\papereightyeighth,\paperninetysecond }\\ 
\emph{Both}&1&search&1&\cite{ \papersixth }\\ 
\emph{Hybrid}&1&search&1&\cite{ \paperseventh }\\ 
\emph{White Box}&15&search&12&\cite{ \paperthird,\paperfourth,\papertenth,\papertwentythird,\papertwentyfifth,\paperthirtyeighth,\paperthirtyninth,\paperfortyfifth,\paperfiftieth,\paperseventieth,\paperseventyfirst,\papereightyfirst }\\ 
&&others&3&\cite{ \paperthirtyfourth,\papersixtyfourth,\paperninetyeighth }\\ 
\midrule 
Total &62& \\ 
\bottomrule 
\end{tabular} 

	\begin{spacing}{0.8}
		\raggedright \footnotesize Note that we only considers the papers which contribute to propose new approach for automated testing of REST APIs.
	\end{spacing}
	\label{tab:rq5}
\end{table}

Based on our survey,
there exist two main types of testing techniques to automate testing of REST APIs, i.e., black-box and white-box.
Regarding black-box testing, it enables to verify systems' behaviors with exposed endpoints, e.g., checking responses of an HTTP request.
The verification and validation of REST APIs with black-box technique are mainly driven by exploiting the API specification and returned responses. 
Based on the results of this survey (see Table~\ref{tab:rq5}), \preOfBB of the existing approaches are in the context of black-box testing.

Regarding the black-box fuzzers, \bBOXRT~\cite{\paperfifth} aims at testing of REST APIs in terms of its robustness. 
\evo applies search-based techniques by defining fitness function with black-box heuristics in a black-box mode~\cite{\papersixth}, such as status code coverage and faulty status code (i.e., 500), to generate tests. 
There are also a set of property-based testing approaches that identify properties of REST APIs as test oracle problems, such as \QuickRest~\cite{\papersixteenth} and \Schemathesis~\cite{\paperfiftyfifth}. % \andrea{need ref}.
For instance, in terms of the API specification, \QuickRest and \Schemathesis both identify consistency with the specification as the properties.
\Schemathesis also explores semantic properties of REST APIs, such as \texttt{GET} should fail after an unsuccessful \texttt{POST} when they perform on the same resource.
Moreover, dependencies of REST APIs are studied for REST API testing.
For instance, 
\RESTler~\cite{\papertwentyfirst} infers and handles dependency among endpoints to generate effective tests based on the API specification and runtime returned responses.
\RestTest generates tests by exploring inter-parameter dependencies of REST APIs.
With \RestTestGen~\cite{\paperfirst,\paperseventeenth}, \textit{Operation Dependency Graph} is proposed  to capture data dependencies among operations of a REST API.
The graph is initially built based on its OpenAPI schema then further extended at runtime.
\morest~\cite{\papersixtythird} formalized a property graph to construct relations of operations and object schema, and such a graph could be derived based on the API schema then dynamically updated based on responses at runtime.
Within a specified time budget, test cases could be initially generated by traversing the graph, and any update of the graph would result in new tests.
Furthermore, 
\RestCT~\cite{\paperfortythird} is a combinational approach, integrating two phases that facilitate generating orders of operations then concertizing input parameters of operations.
The orders are generated based on HTTP action semantics with greedy algorithm, e.g., for a specific resource, its \texttt{GET} operation should not appear before its \texttt{POST} and after its \texttt{DELETE}.
The concrete values of input parameters could be produced in various ways, such as random, previous responses, specified examples or inter-parameter dependencies.
In the context of black-box testing, Cheh and Chen~\cite{\papersixtyfirst} enabled a semi-automatic approach to identify security issues in the API schema.
Navas~\cite{\papereightysixth} proposed an approach to infer and validate the API schema, such as misnamed and duplicated elements in response schema.

Compared to black-box testing, white-box techniques could enable testing of internal behaviors of the REST APIs with additional metrics relating to source code, such as code coverage.
For example, \Pythia~\cite{\papersixtyfourth} employs code coverage heuristics to guide test generation. 
The code coverage could be collected by pre-configuring locations of basic blocks in its implementation with static analysis.
\evo is a fuzzer which has a white-box mode, and the white-box testing is enabled by code instrumentation for JVM~\cite{arcuri2019restful,arcuri2018test} and NodeJS programs~\cite{js2022} developed in the fuzzer.
The code instrumentation allows to identify the code to cover and collect code coverage at runtime.
With such info, \evo defined white-box heuristics and applied search-based techniques to fuzzing REST APIs.
With this SLR, we found that all other white-box testing techniques are built on the top of \evo platform (i.e., 
\cite{\paperthird,\paperfourth,\papertenth,\papertwentythird,\papertwentyfifth,\paperthirtyfourth,\paperthirtyeighth,\paperthirtyninth,\paperfortyfifth,\papereightieth,\papereightyfirst}) with new algorithms and new techniques.

There also exist hybrid approaches which combine black-box and white-box~\cite{\paperseventh}.
For instance, Martin-Lopez~\etal~\cite{\paperseventh} proposed a solution by integrating two fuzzers (i.e., \RestTest and \evo).
A motivation of having such a hybrid approach as described in the paper is that inputs of a REST API are restricted with constraints, and the constraints are not specified in the API schema.
Without the info of constraints, generating successful requests (i.e., receiving a response with 2xx status) is not trivial.
However, white-box testing approach could tackle this problem by identifying the constraints with source code, such as \evo~\cite{\paperfiftieth,\paperseventieth}.
In \cite{\paperseventh}, \RESTest is firstly employed to generate tests in the context of black-box testing, then the generated tests would be regarded as seeds that \evo starts with.
With the four selected REST APIs, the hybrid approach achieved the best results compared to isolated black-box and white-box solutions.

%----------------------------------------------------------
\subsubsection{Search-based Testing}

Search-based testing aims at solving software testing problems with metaheuristic search techniques, such as Genetic Algorithms and Swarm Algorithms.
In order to enable the search techniques, it needs to reformulate the addressed testing problem as a search problem.

\evo is a search-based fuzzer that reformulates test case generation as search problems supporting both black-box and white-box mode.
The black-box mode is achieved with \textit{Random} strategies with black-box heuristics, such as coverage of operations and status code.
The white-box mode is enabled with novel techniques (e.g., code instrumentation~\cite{arcuri2019restful}, testability transformations~\cite{arcuri2021tt,arcuri2020testability} and SQL handling~\cite{arcuri2019sql, arcuri2020handling}) which allows to define white-box heuristics as part of fitness function.
For instance, code instrumentation enables identification of lines and branches to cover as testing targets and collecting such coverage at runtime, and testability transformations provide better guidelines to search for maximizing the testing targets.
Many independent objectives algorithm (MIO) is the default algorithm in white-box mode of \evo.
MIO is an evolutionary algorithm developed specific to white-box system test generation, and the algorithm is inspired by (1+1)EA~\cite{DJW02} that contains sampling and mutation operators.
Its effectiveness to test REST APIs has been demonstrated in several papers~\cite{arcuri2018test,arcuri2019restful,arcuri2021enhancing}. 
To be specialized for Web API testing and REST API domain, MIO is further extended with adaptive hypermutation~\cite{zhang2021adaptive}, smart sampling~\cite{arcuri2019restful} and resource-based techniques~\cite{zhang2021resource,zhang2019resource}.

Genetic Algorithms (GAs) were also enabled for REST API testing.
For instance,
the Whole Test Suite~\cite{GoA_TSE12} employed genetic algorithm to enable automated test suite generation. 
With the GA, the approach evolved the test generation with mutation and crossover operators, and the fitness function is defined with an overall single objective using white-box heuristics. 
Instead of single-objective optimization, 
Many-Objective Sorting Algorithm (MOSA) extends NSGA-II~\cite{deb2002fast} for handling test generations with respects to maximizing many testing targets as many-objective optimization.
WTS and MOSA have been integrated into \evo which allows to apply them to tackle REST API testing~\cite{arcuri2018test}.  
With the \evo platform,
Stallenberg~\etal~\cite{\papertenth} employed Agglomerative Hierarchical Clustering (AHC) to identify patterns of tests, then extended MOSA (named LT-MOSA) to generate tests with the patterns.
Aside from test generation, Liu and Chen~\cite{\papereightyfifth} employed genetic algorithm to optimize test data of REST APIs in the context of mutation testing.

Moreover, Swarm Algorithms were also investigated.
Sahin~\cite{\paperfourth} proposed a Discrete Dynamic Artificial Bee Colony with Hyper-Scout (DABC-HS) algorithm which was introduced to address shortcomings of a basic Artificial Bee Colony (ABC) for REST API testing.
Furthermore, a Greedy Algorithm was applied in \RestCT to produce operation sequences~\cite{\paperfortythird}.

%----------------------------------------------------------
\subsubsection{Property-based Testing}

Property-based testing is an approach which validates and verifies the SUT based on identified properties that should always abide by the SUT.
In the context of REST API testing, one type of properties could be identified based on schema, such as check if the responses during testing conform to the schema~\cite{\papersixteenth,\paperfiftyfifth}.
In addition, as REST, it defines a set of guidelines on how to access resources.
Then, states of resources in REST API are captured as properties of REST APIs.
For instance, \QuickRest~\cite{\papersixteenth} defined stateful properties of REST API with documented responses for producing a stateful sequence of operations.
Seijas~\etal~\cite{\paperthirtieth} generates finite state machines to construct changes of states of resources, then employed Quviq QuickCheck for enabling property-based testing of REST APIs.
Chakrabarti and Rodriquez~\cite{\paperthirtyfirst} examined connectedness of resources in testing REST APIs.

There also exist some work for exploring semantic of REST APIs.
For instance, 
in~\Schemathesis, Hatfield-Dodds and Dygalo~\cite{\paperfiftyfifth} derived structural and semantic properties of REST APIs relating to constraints, security and performances.
Metamorphic testing was also enabled in REST API testing that identified metamorphic relations of the REST API based on their semantics~\cite{\papertwentyseventh,\paperseventyeighth,\paperninetieth}.
 %or security aspects~\cite{\paperninetysixth}.

%----------------------------------------------------------
\subsubsection{Model-based Testing}

Model-based testing is to employ models to perform testing.
The models are typically constructed manually before the testing for depicting the SUT, such as states, behaviors.
For instance, Vu~\etal~\cite{\papertwentieth,\papereightieth,\paperninetyfourth} proposed a model-based testing approach which uses an $\varepsilon$-NFA to construct hypermedia behavior of the REST API.
Fertig and Braun~\cite{\papertwentyninth} developed a Domain Specific Language (DSL) for constructing models of REST APIs, then defined a set of test templates for test generations with such models. 
%\andrea{this sentence is left incomplete}

In~\cite{\paperseventyninth}, Liu~\etal proposed a model-based approach for verifying REST service architecture using colored Petri Nets (CPN).
Five REST feature constraints are defined with CPN models.
Such constraint models allow a verification of architecture models using simulation.
Pinheiro~\etal~\cite{\paperfortysixth} constructed behaviors of REST API with UML state machines.

Moreover, the models could also be used to construct the tests. 
Ed-douibi~\etal~\cite{\papertwentysixth} defined a metamodel for formalizing test suites of REST APIs.
In the approach, the authors proposed a set of rules (e.g., infer parameter values based on examples, generate faulty test cases with invalid inputs) to generate the test suite model based on the OpenAPI.
Then the generated test suite model is further converted to execute code (e.g., JUnit) for performing the requests against the SUT.

%----------------------------------------------------------
%\subsubsection{Formal Verification}

%----------------------------------------------------------
%\subsubsection{Test Data and Test Structure}

%----------------------------------------------------------
\subsubsection{Others}

Graphs were constructed in several works for capturing structures and relationships of REST APIs.
Then, for example Breadth-First Search (BFS) can be employed for generating tests based on such graph~\cite{\papertwentyfirst}.

%TTCN-3~\cite{\papertwelfth}
% might discuss taint analysis

Gazzola~\etal~\cite{\papereightyeighth} proposed an approach \ExVivoMicroTest which facilitates regression test generation by recording service interactions at runtime. 
Godefroid~\etal~\cite{\paperthirtysixth} proposed an approach for detecting faults due to changes made among different versions of RESTful web services.
Such faults could be identified by comparing behavior of different versions with the same inputs generated by a fuzzer \Restler.

Takeda~\etal~\cite{\paperthirtyfourth} developed test case extraction approach which aims at identifying impacts of migration of APIs to microservices from monolithic architecture.
The impacts are derived by analyzing source code in the context of white-box testing.

\begin{result}
	{\bf RQ5}:
	Both white-box and black-box techniques were investigated for tackling testing of REST API, and  most of existing approaches (i.e., \preOfBB) are black-box.  
	With black-box testing, property-based and model-based testing were widely employed that identify characteristics of the REST APIs (such as properties and behavior) to facilitate automated testing.
	White-box testing of REST APIs was mainly addressed by search-based techniques.
	%with \evo by taking its advantage of automatically collecting white-box heuristics using code instrumentation.
\end{result}

%----------------------------------------------------------
\subsection{\rqBTestKind}
\label{subsec:testingkind}

To find out what kinds of testing have been conducted by the papers and what are their frequencies, we extracted relevant data and analyzed them. The results of our findings are shown in Table~\ref{tab:rq6}.
Not all the papers necessarily conduct only one testing type and they might cover more than one of them (\eg unit testing, integration testing and acceptance testing are covered by~\cite{\papereightyfourth}).
%Detected testing types we have found are as follows:
With this SLR, we discovered \testingTypes testing types in the context of REST API testing which had been investigated by the papers.
More details about these types are discussed as follows:

\begin{itemize}
	
	\item \textbf{System Testing}
	
	System testing is a type of testing that verifies a software product's integration and completion. A system test's objective is to gauge how well the system requirements are met from beginning to end. Most of the papers (i.e., \System out of \totalPapers) we found are either focusing solely on this type (\eg \cite{\papereighth}) or are conducting it along with other types.
	As the REST is a guideline to build the web services and current techniques (e.g., OpenAPI) enables necessary info to perform system testing (e.g., make the request to REST API), it is expected that the system testing is the most addressed problem in REST API testing.

	%(\eg EvoMaster which also can be used for regression testing, according to its website~\cite{EvoMaster}).
	
	% Man: move it below based on the number
	%\item \textbf{Regression Testing}
	%
	%Regression testing is the process of ensuring that altered software continues to function as intended~\cite{engstrom2008empirical}. For example, in order to automatically find breaking changes across API versions, differential regression testing for REST APIs is presented in~\cite{\paperthirtysixth}. In this study, \Regression papers
	%%have covered
	%tackle specific regression testing problems for REST API.
	%Moreover,
	%%some
	%test cases produced by fuzzer could also be used for regression testing. For example, \evo automates system testing but the test cases generated by this tool can be used for conducting regression testing.
	
	\item \textbf{Security Testing}
	
	The fundamental objective of security testing is to determine the system's risks and assess any potential vulnerabilities, so that threats can be confronted and the system can continue to operate without being compromised. As an example, the study by Cheh and Chen uses the standardized OpenAPI specification as an input and suggests a semi-automatic method to deduce different significant details regarding the security flaws in that API definition~\cite{\papersixtyfirst}. Security testing of REST APIs is of great importance. As it is mentioned in Section~\ref{sec:intro}, many large enterprises rely on REST APIs. However, there have reportedly been a number of incidents involving web API security in recent years. The top three were denial of service attacks (19\%), bot/scraping (20\%), and vulnerabilities (54\%) followed by authentication problems (46\%). These flaws continue to exist until a hacker finds and takes advantage of them, which can lead to data loss, account abuse, or service interruption~\cite{SaltSecurity}.
	
	\item \textbf{Integration Testing}
	
	This stage examines whether collections of components function as expected by the technical system design or specification. We found some papers focusing on this kind of testing, such as the study by Vu \etal~\cite{\papereightieth} which is aimed at automation of integration testing with the focus on hypermedia testing.
	
	\item \textbf{Unit Testing}
	
	Unit testing is the process of testing a single units, small specialized section of code written by a developer. MockRest~\cite{\paperseventysixth} is one tool focusing on unit testing by proposing a mock framework to help developers get a consistent response while the real REST API is down.

	\item \textbf{Regression Testing}
	
	Regression testing is the process of ensuring that altered software continues to function as intended~\cite{engstrom2008empirical}. For example, in order to automatically find breaking changes across API versions, differential regression testing for REST APIs is presented in~\cite{\paperthirtysixth}. In this study, \Regression papers
	%have covered
	tackle specific regression testing problems for REST API.
	Moreover,
	%some
	test cases produced by a fuzzer could also be used for regression testing. For example, \evo automates system testing but the test cases generated by this tool can be used for conducting regression testing.
	
	\item \textbf{Robustness Testing}
	
	Robustness testing aims to identify the extent to which a particular system or component can continue to operate properly in the presence of erroneous inputs or demanding environmental circumstances~\cite{IEEEStandardSoftwareTerminology}.
	Fuzzers that aim at finding faults in the APIs might send invalid inputs on purpose to check if the API correctly returns an error message.
	One work focusing on robustness testing is \cite{\paperfifth}, which performed this type of test over REST services based on the constraint information expressed in their OpenAPI specification.

	\item \textbf{Architecture Design Testing}
	
	The only paper which we found conducting this kind of testing is the study by Liu \etal~\cite{\paperseventyninth}.
	This paper is aimed at enhancing system design's quality by verifying whether an API conforms to the REST architecture constraints described in Section~\ref{subsec:http}.

	\item \textbf{Acceptance Testing}
	
	Acceptance testing is a formal testing procedure used to ascertain whether a system satisfies its acceptance criteria and to give the client the option of accepting the system or not. The study by Besso \etal is the only paper which covers this type of testing by conducting it against web service choreographies~\cite{\papereightyfourth}.
	
\end{itemize}

\begin{table}
	\small
	\centering
	\caption{Different testing types and their frequencies among the papers}
	\begin{tabular}{ l l p{0.5\textwidth}} 
\toprule 
Testing Type & \# & Papers \\ 
\midrule 
System & 72 & \cite{\paperfirst,\papersecond,\paperthird,\paperfourth,\paperfifth,\papersixth,\paperseventh,\papereighth,\paperninth,\papertenth,\papereleventh,\papertwelfth,\paperthirteenth,\paperfifteenth,\papersixteenth,\paperseventeenth,\papereighteenth,\papernineteenth,\papertwentieth,\papertwentyfirst,\papertwentysecond,\papertwentythird,\papertwentyfourth,\papertwentyfifth,\papertwentysixth,\papertwentyseventh,\papertwentyninth,\paperthirtieth,\paperthirtyfirst,\paperthirtysecond,\paperthirtythird,\paperthirtyfourth,\paperthirtyeighth,\paperthirtyninth,\paperfortieth,\paperfortyfirst,\paperfortysecond,\paperfortythird,\paperfortyfourth,\paperfortyfifth,\paperfortysixth,\paperfortyeighth,\paperfiftieth,\paperfiftyfirst,\paperfiftysecond,\paperfiftythird,\paperfiftyfourth,\paperfiftyfifth,\paperfiftysixth,\papersixtieth,\papersixtysecond,\papersixtythird,\papersixtyfourth,\papersixtyfifth,\papersixtysixth,\papersixtyseventh,\papersixtyninth,\paperseventieth,\paperseventyfirst,\paperseventysecond,\paperseventythird,\paperseventyfourth,\paperseventyeighth,\papereightyfirst,\papereightythird,\papereightyfifth,\papereightysixth,\paperninetieth,\paperninetysecond,\paperninetyfourth,\paperninetyseventh,\paperninetyeighth}\\ 
Security & 8 & \cite{\paperfourteenth,\papertwentyeighth,\paperthirtyseventh,\paperfortyninth,\papersixtyfirst,\papersixtyeighth,\paperseventyfifth,\paperseventyseventh}\\ 
Integration & 8 & \cite{\papertwentieth,\paperthirtyfifth,\paperfiftyseventh,\paperfiftyeighth,\papereightieth,\papereightysecond,\papereightyfourth,\paperninetyfirst}\\ 
Unit & 5 & \cite{\paperthirtyfourth,\paperseventysixth,\papereightysecond,\papereightyfourth,\paperninetyfirst}\\ 
Robustness & 3 & \cite{\paperfirst,\paperfifth,\paperseventeenth}\\ 
Regression & 2 & \cite{\paperthirtysixth,\papereightyeighth}\\ 
Architecture & 1 & \cite{\paperseventyninth}\\ 
Acceptance & 1 & \cite{\papereightyfourth}\\ 
\bottomrule 
\end{tabular} 

	\label{tab:rq6}
\end{table}

As it is shown in Table~\ref{tab:rq6}, the frequency of papers which focus on system testing was much higher than those papers that necessarily require accessing to software components (\eg integration test) or source code (\eg unit test). As mentioned in Section~\ref{subsec:http}, REST is a high-level design guideline, so it is understandable that most of the papers (\ie~\System) were focusing on system testing which is performed to determine a complete software system is working properly.

\begin{result}
	{\bf RQ6}:
	Existing studies in REST API testing covered \testingTypes different testing types.
	Most of the studies (i.e., \System out of \totalPapers) referred to system testing of REST API. 
%	Integration and security testing were the second most common types by being ad
%Existing studies in REST API testing covered \testingTypes different testing types in which system testing was the most common one by covering most of the papers (\ie~\System out of \totalPapers).

\end{result}

%----------------------------------------------------------
\subsection{\rqBCaseStudy}
\label{subsec:artifacts}

\begin{figure}[!th]
	\centering
	%	\hspace*{0.4in}
	\includegraphics[scale=0.6]{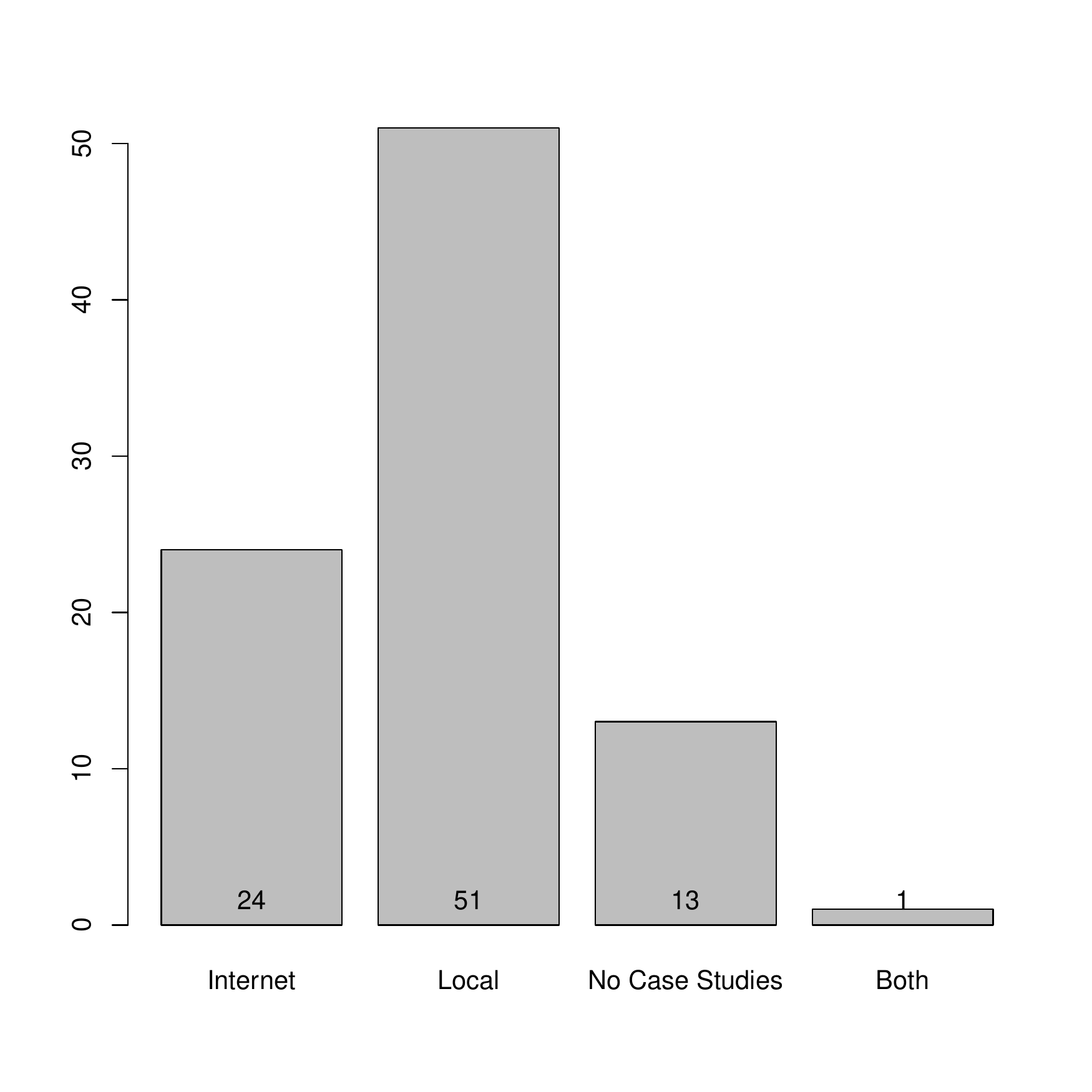}
	\vspace{-1.5\baselineskip}
	\caption{Location of Case Studies Used for Empirical Evaluations and Their Frequencies}
	\label{fig:rq7-histogram}
\end{figure}

In this section, we investigated what kind of artifacts researchers have used as case studies to empirically evaluate the effectiveness of their proposed approaches.
Sound empirical evidence is needed to demonstrate the usefulness of a novel technique.
The larger and more variegated a case study is, the more likely it will be that results would generalize to other systems.

The results of our findings, including different categories of case studies and their frequency among the papers are displayed in Figure~\ref{fig:rq7-histogram}.
When dealing with experiments on testing Web APIs, there are two main types of artifacts:
(1) APIs run on a local machine;
and (2) existing APIs available on the internet (e.g., ProgrammableWeb~\cite{ProgrammableWeb} currently lists more than 20 000 APIs).

%We classified the case studies based on availability (i.e., up and running on internet or local), accessibility (i.e., open source or closed source) and its application context (i.e., industrial, real\man{might need a type for case studies, such as features-service, proxyprint} or artificial).
%In total, we classified them into \artefactTypes types, i.e., \man{names of the categories might not be consistent, eg, might need to include three aspects for each. In addition, ``industrial case studies'' might be an important aspect to consider, but now it is a bit missing in the results.}.

For the former type (i.e., APIs run on a local machine), those are typically open-source projects, hosted on open-source repositories such as GitHub.
Local open-source project is the most common type of case studies used by \LocalOpenSource of the papers.
For instance, case studies from \textit{EMB} repository\footnote{https://github.com/EMResearch/EMB} which includes a set of open source APIs are utilized by \numberOfPapersUsingEmb papers as case study~\papersUsingEmb.
There are cases in which local closed-source APIs are used for this kind of experiments, but it is a less common occurrence (\ie \LocalClosedSource), as it typically requires academia-industry collaborations on joined research projects~\cite{garousi2019characterizing}.
There are also papers which take advantage of both open-source and closed-source APIs by running them locally.
This group consists of \LocalOpenSourceandClosedSource of the papers.
In addition, several cases are witnessed in which some artificial example APIs are built by researchers to conduct empirical studies.
These artificial artifacts are utilized by \LocalArtificial of the papers.

The latter type, which include \Internet of the case studies, are typically industrial APIs providing paid services on internet, or free services from different government agencies.
Some might be open-source, but those are a small minority.
On this kind of APIs, typically only black-box techniques are investigated, as white-box techniques require access to the source code.
For example, \RestTestGen~\cite{\paperfirst,\paperseventeenth} used 87 RESTful APIs available on internet for its empirical studies.
However, one major drawback of using APIs on the internet is that experiments might not be \emph{repeatable}, as APIs on the internet might change or disappear without any previous notice.
There is also the case of \cite{\papereightysixth}, which uses case studies hosted on the internet along with artifacts developed by the researchers.

Finally, there are a considerable number of papers which did not perform any empirical evaluations. 
\NoCaseStudies of papers, such as the \PINumber papers which their main contribution is of type ``\PI'' did not conduct any empirical evaluations.
There are also papers which are not of type ``\PI'', but did not make use of any case studies for empirical evaluations, such as the study by Pinheiro \etal ~\cite{\paperfortysixth}.

\begin{result}
	{\bf RQ7}: The most common group of case studies are local open-source projetcs and APIs on the Internet which are used by \LocalOpenSource and \Internet of the papers, respectively.
	We also discovered that \NoCaseStudies of the papers do not use any case studies.
\end{result}

%%%%%%%%%%%%%%%%%%%%%%%%%%%%%%%%%%%%%%%%%%%%%%%%%%%%%%%%%%%%%%%%%%%%%%%%%%%%
\section{Tools for Testing RESTful APIs}
\label{sec:tools}

%----------------------------------------------------------
\subsection{\rqCOpenSou}
\label{subsec:openSource}

\begin{table}[!ht]
	\small
	\centering
	\caption{
		%Papers with open source tool or replication package
		Information of open-source research tools have been developed for REST API testing
	}
	%	\vspace{-1\baselineskip}
	\label{tab:Open-source-tools}
	\resizebox{1\textwidth}{!}{
		
\begin{tabular} { l l l l p{0.25\textwidth} p{0.25\textwidth}}\\ 
\toprule 
& Tool & Website & \# & Papers & Programming Language(s) \\ 
\midrule 
1 & EvoMaster & \cite{EvoMaster} & 19 & \cite{\paperthird,\paperfourth,\papersixth,\paperseventh,\papertenth,\papertwentythird,\papertwentyfifth,\paperthirtyeighth,\paperthirtyninth,\paperfortieth,\paperfortyfourth,\paperfortyfifth,\paperfiftieth,\paperfiftyfirst,\paperseventieth,\paperseventyfirst,\paperseventythird,\papereightyfirst,\paperninetyeighth} & Java, Kotlin, C\#, TypeScript, Javascrip\\ 

2 & RESTest & \cite{RESTest} & 8 & \cite{\paperseventh,\paperninth,\papereleventh,\paperfortieth,\paperfortyfirst,\paperfortyfourth,\paperfiftythird,\paperninetysecond} & Java\\ 

3 & Restler & \cite{Restler} & 8 & \cite{\paperninth,\paperfourteenth,\papereighteenth,\papertwentyfirst,\paperthirtysixth,\paperfortieth,\paperfortyfourth,\papersixtysixth} & Python, F\#\\ 

4 & RestTestGen & \cite{RestTestGenV2} & 5 & \cite{\paperfirst,\paperninth,\paperseventeenth,\paperfortieth,\paperfortyfourth} & Java\\ 

5 & bBOXRT & \cite{bBOXRT} & 4 & \cite{\paperfifth,\paperninth,\paperfortieth,\paperfortyfourth} & Java\\ 

6 & Schemathesis & \cite{Schemathesis} & 3 & \cite{\paperfortieth,\paperfortyfourth,\paperfiftyfifth} & Python\\ 

7 & RestCT & \cite{RestCT} & 2 & \cite{\paperfortieth,\paperfortythird} & Python\\ 

8 & Jsongen & \cite{Jsongen} & 2 & \cite{\papersixtieth,\papereightythird} & Erlang\\ 

9 & RESTApiTester & \cite{RESTApiTester} & 1 & \cite{\papersecond} & Java\\ 

10 & DeepLearningBasedTool & \cite{DeepLearningBasedTool} & 1 & \cite{\papereighth} & Python\\ 

11 & Api Tester & \cite{ApiTester} & 1 & \cite{\papertwentysixth} & Java\\ 

12 & Restats & \cite{Restats} & 1 & \cite{\paperthirtythird} & Python\\ 

13 & Gadolinium & \cite{Gadolinium} & 1 & \cite{\paperfortysecond} & TypeScript, JavaScript\\ 

14 & Hsuan-Fuzz & \cite{Hsuan-Fuzz} & 1 & \cite{\paperfiftysecond} & Go\\ 

15 & Instance Identification & \cite{InstanceIdentification} & 1 & \cite{\paperfiftyseventh} & Go\\ 

16 & ExVivoMicroTest & \cite{ExVivoMicroTest} & 1 & \cite{\papereightyeighth} & Python\\ 

\bottomrule 
\end{tabular} 
}
\end{table}

Typically, a scientific article does not have the space to fully describe in details all the low level details of a new presented technique.
Releasing the implementation of a new research prototype as open-source not only helps in this aspect, but it is also helpful to enable replicated studies (as re-implementing everything from just an article description can be a major engineering task).
Furthermore, open-source tools, if maintained and properly engineered, can be used by practitioners to reduce the gap between academic research and industrial practice.

Studies that have their tool released as open-source, or use existing open-source tools for comparisons, are listed in Table~\ref{tab:Open-source-tools}, along with the programming language(s) the tools are implemented with.
Note that, in some cases, studies extend on existing open-source tools, but no info is provided on whether the extension is available as open-source.
Also, it might happen that a tool is released as open-source only \emph{after} a scientific article is published.
We have included these cases as well, but due to the lack of precise information (e.g., URL links might not be explicitly stated in the articles) we might have missed some.

Overall, it can be observed that \NumberOfOpenSourceTools tools have either been compared or released by \NumberOfPapersWithOpenSourceTools papers.
%\MostUsedOpenSourceTool 
\evo is the one which the highest number of papers.
\NumberOfPapersUsingEvoMaster papers are either presenting it or proposing a new technique which is integrated into it or using it for comparison with other tools.
The second most common tools are \RestTest and \RESTler, used in \NumberOfPapersUsingRestler papers.

It can be seen that \MostUsedProgrammingLanguageByOpenSourceTool and \SecondMostUsedProgrammingLanguageByOpenSourceTool have been the most common programming languages,  used by \NumberOfMostUsedProgrammingLanguageByOpenSourceTool open-source tools each.

\begin{result}
	{\bf RQ8}: 
	%Several  studies have published their proposed tools as open source, including 16 tools.
	%\evo and RESTest are the most frequent tools among the analyzed studies.
	\NumberOfOpenSourceTools  research tools have been released as open-source based on research outcomes of REST API testing.
	\evo is the most used tool which have been studied, extended and compared with other tools, and \restest and \RESTler are the second most ones.  
	\MostUsedProgrammingLanguageByOpenSourceTool and \SecondMostUsedProgrammingLanguageByOpenSourceTool are the most applied programming language for developing the research tools.
\end{result}

%----------------------------------------------------------
\subsection{\rqCNonResTool}

REST APIs are widely used in industry, and their testing is a concrete issue that needs to be addressed, as it has practical value.
Therefore, besides the work from the research community, it is not surprising to see effort from engineers and practitioners in industry to address this problem.
Our goal here is not to survey all the non-research work done on the topic, but rather to analyze what academics found important and relevant enough to cite and use in their studies.
Note that, with the term ``non-research'' we loosely mean all the software tools and libraries developed by practitioners in industry, without any published scientific article (as far as we know) describing them.

The aim of this research question is to find out what are the non-research tools and libraries that have been used in our surveyed studies.
Those should be either specific to test REST APIs (e.g., API fuzzers), or can be used for testing REST APIs (e.g., code coverage tools).
The tools we found are listed in Table~\ref{tab:rq9}.
We found a larger number of non-research tools being used or compared in the papers, but they all were not relevant to software testing or REST APIs domains.
For example, \textit{WSFuzzer}\footnote{https://sourceforge.net/projects/wsfuzzer/} is a tool used by one of the papers~\cite{\papersixtyninth}, but it is aiming at SOAP APIs, as the paper covers this type of APIs as well.
%
%\andreaAddressed{this needs some better restructuring. eg libraries grouped in same paragraph vs fuzzers}
%
With this SLR, we found two kinds of non-research tools have been used in REST API testing, i.e., \textit{library}, and \textit{toolkit}.
%\man{To Andrea: fuzzer might not be applicable to postman, then how about toolkit? with the current description,  the framework might be a bit overlap with toolkit, not sure if we need it. In addition,
	%description column in the table 8 is not so clear, might show the kind of tool instead of the description. if the update is needed, @Amid might do it.}.

There exist two libraries which have been used in testing of REST API. 
Being utilized by \restassuredNumber papers, \textit{RestAssured}~\cite{RestAssured} is the most 
%common 
used non-research tool
%It is a library that offers a domain-specific language (DSL) for creating tests for RESTful APIs in Java.
which facilitates doing HTTP calls against REST API and validating responses.
\textit{RestAssured} is mainly applied in tests generated by fuzzers, such as \evo~\cite{arcuri2019restful} and \restest~\cite{\paperseventh}.
\textit{Swagger Schema Validator}~\cite{swaggerschemavalidator} is another library which validates JSON objects against Swagger 2 which is utilized by response validation oracle to assess the automated test case generation~\cite{\paperfirst}.

%Similarly, \textit{Swagger-conform} aims to find any places where the REST API fails to adhere to its own schema specified in Swagger/OpenAPI~\cite{swaggerconform}. 
%This tool is evaluated by one paper~\cite{\paperfiftyfifth}.

% move to toolkit
%We also found two testing frameworks.
%\textit{Fuzz-lightyear}~\cite{fuzzlightyear} is a framework for stateful fuzzing.
%It maintains the state between requests, allowing us to put together a request sequence, design it to simulate a malicious attack vector, and use it to notify of unexpected success.
%\textit{Go-fuzz}~\cite{gofuzz} is another framework to enable fuzzing in applications written in Go language.

%Rest of the tools are either used or compared by only one paper.
There are variety of 
%other tools
toolkit for testing of REST API.
For instance, 
%, from 
\textit{Postman}\cite{postman} 
%which is a client tool for conducting manual tests on APIs to fuzzer tools.
is a platform for building and testing APIs, such as test execution and result validation.
For REST API testing, tests can be specified as \textit{Postman} format, and such tests can be created manually or automatically by fuzzers.
For instance, \evo supports \textit{Postman} tests as seed for automated test case generation~\cite{\paperseventh}.
% move framework here
\textit{Fuzz-lightyear}~\cite{fuzzlightyear} is a framework for stateful fuzzing.
It maintains the state between requests, allowing us to put together a request sequence, design it to simulate a malicious attack vector, and use it to notify of unexpected success.
\textit{Go-fuzz}~\cite{gofuzz} is also a framework to enable fuzzing in applications written in Go language.
\textit{Burp Suite}~\cite{burpsuite} is a commercial fuzzer for security testing that is utilized by \burpsuiteNumber papers.
\textit{TnT-Fuzzer}~\cite{tntfuzzer} is an open-source tool for testing robustness which is taken advantage of for evaluation by \tntfuzzerNumber papers~\cite{\paperfiftyfifth,\papersixtyfifth}.
\textit{SoapUI}~\cite{soapui} is an application to perform automated end-to-end tests on a variety of web-services, including REST APIs to test their performance and security.
\textit{Tcases}~\cite{tcases} is another tool that performs black-box model-based testing based on OpenAPI specifications.
\textit{SpotBugs}~\cite{spotbugs} is a tool which enables static analysis of codes written in Java.
\textit{Dredd}~\cite{dredd} is an API testing tool which conducts sample-value-based testing technique and validates responses based on status codes, headers and body payloads.
\textit{Autorize}~\cite{autorize} is another tool focusing on security which detects authorization enforcement within a REST API.
\textit{ZAP}~\cite{zap} is a web application scanner tool focusing on security testing by performing penetration testing.
%\textit{Got-Swag}~\cite{gotswag} is a tool for random testing of APIs based on Swagger.
Rest of the tools are fuzzers which depend on OpenAPI/Swagger schema.

\begin{table}[!ht]
	\small
	\centering
	\caption{Information of non research tools which have been applied in REST API testing}
	\label{tab:rq9}
	\resizebox{1\textwidth}{!}{
		\begin{tabular}{ l p{0.2\textwidth} l p{0.4\textwidth} l p{0.15\textwidth}} 
\toprule 
& Tool & Website & Description & \# & Papers \\ 
\midrule 
1 & RestAssured & \cite{RestAssured} & DSL for writing tests in Java & 12 & \cite{\paperseventh,\papereleventh,\papertwentythird,\paperthirtyeighth,\paperthirtyninth,\paperfiftieth,\paperseventieth,\paperseventyfirst,\paperseventyfifth,\papereightyfirst,\paperninetysecond,\paperninetyeighth}\\ 

2 & Swagger Schema Validator & \cite{swaggerschemavalidator} & Validates JSON objects against Swagger 2 & 2 & \cite{\paperfirst,\paperseventeenth}\\ 

3 & Postman & \cite{postman} & Client tool for testing APIs manually & 2 & \cite{\paperseventh,\papereleventh}\\ 

4 & Burp Suite & \cite{burpsuite} & Commercial security testing fuzzer & 2 & \cite{\paperthirtythird,\papersixtyfifth}\\ 

5 & SoapUI & \cite{soapui} & Web-service testing application & 2 & \cite{\paperfortysecond,\papereightysixth}\\ 

6 & APIFuzzer & \cite{apifuzzer} & Fuzzer tool based on OpenAPI & 2 & \cite{\paperfortyfourth,\paperfiftyfifth}\\ 

7 & Fuzz-lightyear & \cite{fuzzlightyear} & Stateful fuzzing framework based on OpenAPI & 2 & \cite{\paperfortyninth,\paperfiftyfifth}\\ 

8 & TnT-Fuzzer & \cite{tntfuzzer} & Open-source robustness testing tool & 2 & \cite{\paperfiftyfifth,\papersixtyfifth}\\ 

9 & SpotBugs & \cite{spotbugs} & Static analysis tool for Java code & 1 & \cite{\paperfifth}\\ 

10 & Dredd & \cite{dredd} & Validates responses based on status codes, headers, and body payloads & 1 & \cite{\paperfortyfourth}\\ 

11 & Tcases & \cite{tcases} & Black-box model-based testing based on OpenAPI & 1 & \cite{\paperfortyfourth}\\ 

12 & Autorize & \cite{autorize} & Authorization enforcement detection tool (Extension for Burp Suite) & 1 & \cite{\paperfortyninth}\\ 

13 & Go-fuzz & \cite{gofuzz} & Randomized testing for Go & 1 & \cite{\paperfiftysecond}\\ 

14 & Got-Swag & \cite{gotswag} & Monkey testing for APIs based on Swagger & 1 & \cite{\paperfiftyfifth}\\ 

15 & Cats & \cite{cats} & Fuzzer and negative testing tool based on OpenAPI & 1 & \cite{\paperfiftyfifth}\\ 

16 & Swagger-conform & \cite{swaggerconform} & Tests if API conforms to Swagger schema & 1 & \cite{\paperfiftyfifth}\\ 

17 & Fuzzy-swagger & \cite{fuzzyswagger} & Fuzzer based on Swagger & 1 & \cite{\paperfiftyfifth}\\ 

18 & Swagger Fuzzer & \cite{swaggerfuzzer} & Fuzzer based on Swagger & 1 & \cite{\paperfiftyfifth}\\ 

19 & Fuzzapi & \cite{fuzzapi} & Open-source general-purpose HTTP fuzzer & 1 & \cite{\papersixtyfifth}\\ 

20 & ZAP & \cite{zap} & Penetration Testing Tool For Web Apps & 1 & \cite{\papersixtyninth}\\ 

\bottomrule 
\end{tabular} 
}
\end{table}

\begin{result}
	{\bf RQ9}: 
	\NumberOfNonResearchTools non-research tools 
	%and libraries 
	were identified
	%in the analyzed articles,
	from the selected \totalPapers papers for REST API testing,
	in which RestAssured was the most
	% frequent one by being
	used/compared by \restassuredNumber papers.
\end{result}

%----------------------------------------------------------
\subsection{\rqCFeatures}

To get better understanding of the current state-of-the-art, 
we investigated what features are supported by the research prototypes.
This can also help practitioners to evaluate these prototypes.
To answer this research question, 
we collected features supported by the \NumberOfTools open source tools which we identified (see Section~\ref{subsec:openSource}).
Based on the collected features,
we then divided them into \RQTenCategoriesNumber main categories as 
shown in Table \ref{tab:rq10}.
The identification of features is based on what reported in the published papers of these tools, and their online documentation (if any is available).
At times, the needed information is either unclear of missing.
For example, a tool could state that it generates ``test cases'', but no info seems provided on the language and format of these output test cases in the documentation.

%Based on \ref{tab:rq10},
OpenAPI is supported by most of the tools (i.e., \NumberOfToolsSupportingOpenAPI out of \NumberOfTools) as a supported format of REST API schema.
The tools take the OpenAPI specification as an input to identify endpoints of the REST API.
It has also been taken advantage of alongside other information such as HTTP logs in Restats and visualized data (\ie charts) in Gadolinium~\cite{\paperfortysecond}.

Regarding what the released prototypes output, test cases are the most common one and among them, \textit{JUnit} is the most employed one.
%JUnit is the only output format of Api Tester, however, it is not the only supported output format in EvoMaster and RestTestGen.
Since \evo supports other programming languages (\ie C\# and Javascript) it also supports tests written in these languages (\eg xUnit and Jest).

The capability to test REST APIs which needs the client to be authenticated/authorized is a challenge that is handled by \NumberOfToolsSupportingAuth of the tools.
Another supported feature is clearing the database after each test run which resets the database state.
This feature is of great importance, as each test case has to be independent from each other.
This feature is only supported by \evo.
\evo is also the only tool which supports \textit{Automated Code Instrumentation} as it conducts white-box testing and needs to insert probes into the source code of the SUT to collect code coverage during test case generation.

The features we detected are not only limited to Table~\ref{tab:rq10}.
For example, installers for different operating systems are available for \evo.
As another example, \restest enables user to modify operations under test.
In other words, it is allowed to exclude some endpoints from being tested.

\begin{table}[!ht]
	\small
	\centering
	\caption{Features which have been supported by open-source research prototypes}
	%	\vspace{-1\baselineskip}
	\label{tab:rq10}
		\resizebox{1.0\textwidth}{!}{
		
\begin{tabular} { l l p{0.6\textwidth}}\\ 
\toprule 
Item & \# & Tool(s) \\ 
\midrule 
\textbf{Input} & & \\ 
OpenAPI  &  12  &  RestTestGen, RESTApiTester, EvoMaster, bBOXRT, RESTest, Restler, Api Tester, Restats, Gadolinium, RestCT, Hsuan-Fuzz, Schemathesis \\ 
I/O Traces  &  2  &  Restats, ExVivoMicroTest \\ 
API Call  &  1  &  DeepLearningBasedTool \\ 
Path dependencies  &  1  &  Hsuan-Fuzz \\ 
JSON schema  &  1  &  Jsongen \\ 
\midrule 
\textbf{Output} & & \\ 
JUnit test case  &  4  &  RestTestGen, EvoMaster, RESTest, Api Tester \\ 
Unspecified test case  &  4  &  RESTApiTester, bBOXRT, Restler, Hsuan-Fuzz \\ 
Python test case  &  2  &  Schemathesis, ExVivoMicroTest \\ 
xUnit test case  &  1  &  EvoMaster \\ 
Jest test case  &  1  &  EvoMaster \\ 
Validity of API call  &  1  &  DeepLearningBasedTool \\ 
Test coverage metrics  &  1  &  Restats \\ 
 Test coverage level  &  1  &  Restats \\ 
OpenAPI  &  1  &  Gadolinium \\ 
Visualized data (charts)  &  1  &  Gadolinium \\ 
Test results (statistics.csv and Swagger)  &  1  &  RestCT \\ 
QuickCheck generator  &  1  &  Jsongen \\ 
\midrule 
\textbf{Authentication Support} & & \\ 
 &  8  &  RestTestGen, EvoMaster, RESTest, Restler, Api Tester, Hsuan-Fuzz, Instance Identification, Jsongen  \\ 
\midrule 
\textbf{Database Reset Support} & & \\ 
 &  1  &  EvoMaster  \\ 
\midrule 
\textbf{Automated Code Instrumentation} & & \\ 
 &  1  &  EvoMaster  \\ 
\bottomrule 
\end{tabular} 

		}
\end{table}

\begin{result}
	{\bf RQ10}: There are a variety of features supported by the released prototypes.We listed them based on \RQTenCategoriesNumber categories. OpenAPI schema as inputs and JUnit test cases as outputs have been the most common ones.
\end{result}

%%%%%%%%%%%%%%%%%%%%%%%%%%%%%%%%%%%%%%%%%%%%%%%%%%%%%%%%%%%%%%%%%%%%%%%%%%%%
\section{Research Challenges}
\label{sec:challenges}

In this section, we discuss 
%the most common 
challenges presented in 
%the analyzed papers.
existing studies of REST API testing.
These could be 
%solved 
addressed
challenges that have been posed by the research questions in the papers  (RQ11), or issues that are still open for future work (RQ12).
%----------------------------------------------------------
\subsection{\rqDAdresChal}
\label{sec:AdresChal}

\begin{table}
	\small
	\centering
	\caption{Common research challenges addressed by the papers}
	
\begin{tabular}{ l l p{0.5\textwidth}} 
\toprule 
Challenge & \# & Papers \\ 
\midrule 
Handling resource dependencies & 8 & \cite{\paperthird,\papertenth,\papertwentyfirst,\papertwentyfifth,\paperthirtyfirst,\paperthirtyfifth,\paperfortyeighth,\papersixtythird} \\ 

Inferring inter-parameter dependencies & 6 & \cite{\paperfortythird,\paperfiftythird,\paperfiftysixth,\papersixtyfifth,\paperninetysecond,\paperninetyseventh} \\ 

Oracle problem & 5 & \cite{\paperfourteenth,\papertwentyseventh,\paperfortyninth,\paperseventyeighth,\paperninetieth} \\ 

Handling database & 3 & \cite{\paperthirtyninth,\paperfortyfifth,\paperseventyfirst} \\ 

White-box heuristics & 3 & \cite{\paperfiftieth,\paperseventieth,\paperninetyeighth} \\ 

Mocking & 2 & \cite{\papertwentysecond,\paperseventysixth} \\ 

Defining coverage criteria & 2 & \cite{\papertwentyfourth,\paperthirtythird} \\ 

Instance identification & 2 & \cite{\paperfiftyseventh,\paperfiftyeighth} \\ 

\bottomrule 
\end{tabular} 

	\begin{spacing}{0.8}
		%		\raggedright \footnotesize Note that we only considers the papers which contribute to propose new approach for automated testing of REST APIs.
	\end{spacing}
	\label{tab:rq11}
\end{table}

We studied the papers to infer the main challenge(s) they have addressed.
Then, we wrote them down, grouped them and searched for the most common ones, which appear in at least more than one article.
Most articles deal with presenting a novel technique (usually a fuzzer) aiming at fault finding.
Here, however, we rather discuss work that focuses on specific identified challenges of testing REST APIs, and not their testing in general.
This can  provide a better understanding of different specific aspects of testing REST APIs.

In Table~\ref{tab:rq11}, we have included the research challenges which have been addressed by at least one paper.
\textit{Handling resource dependency} was the most repeated research challenge among the analyzed articles,
focused on in \handlingresourcedependenciesNumber articles.
This challenge refers to dependencies among resources as they typically exist in the SUTs, and \textit{dependency identification} is used to detect such dependencies based on \textit{REST API Schema}, \textit{Accessed SQL Tables} and \textit{Fitness Feedback}~\cite{\paperthird}.
For example, Stallenberg \etal~\cite{\papertenth} have formed a model which captures, replicates and preserves dependency patterns of API calls in new test cases as breaking them could impede the effectiveness of the test case generation process.

 \emph{Inferring inter-parameter dependencies}
  was the second most frequent addressed challenge (examined in \inferringinterparameterdependenciesNumber articles).
 It refers to the restrictions that web services frequently impose on how two or more input parameters can be combined to create valid calls to the service.
It is frequent that the use of one parameter necessitates or impedes the use of another parameter or set of parameters.
For example, \cite{\paperfiftysixth} has introduced a domain-specific language for the formal specification of dependencies, called Inter-parameter Dependency Language (IDL), and a tool suite for the automated analysis of IDL.

Another common challenge is \textit{Oracle problem}.
Fuzzers can identify faults based on 500 HTTP status code, and mismatches of the responses with the given API schema.
Research has been carried out to define further automated oracles to be able to detect more faults.
This could be based on security rules (e.g.,~\cite{\paperfourteenth,\paperfortyninth}), or metamorphic relations.
When a test execution's expected outcome is complex or unclear, metamorphic testing offers a solution that solves the oracle problem~\cite{chen2020metamorphic,segura2016survey}.
Instead of examining the results of a single program execution, metamorphic testing examines whether many instances of the program being tested satisfy particular requirements known as metamorphic relations.
For example, take into account the following metamorphic relation in Spotify: \textit{regardless of the size of the pagination, two searches for albums with the same query should return the same number of total results}~\cite{\papertwentyseventh}.

%Metamorphic testing is utilized by \metamorphictestingNumber papers to alleviate the oracle problem.
%For instance, Mai \etal have taken advantage of metamorphic testing to alleviate the oracle problem in security testing~\cite{\paperninetysixth}.

The rest of challenges shown in Table~\ref{tab:rq11} are less common, as they were addressed by two-three articles each.
\textit{Handling Databases} is a challenge which was addressed as it has proven that taking database's state into account when generating tests will result in higher code coverage and finding new faults~\cite{\paperthirtyninth,\paperseventyfirst}.

\textit{White-box heuristics} aim at improving code coverage results.
For example, a common issue in SBST is the \textit{flag problem}~\cite{BaS03}, where the branch distance is not able to provide any gradient.
To solve this problem, one method is to use so-called Testability Transformations to change the SUT's source code in a way that enhances the fitness function~\cite{HHH04}.
These techniques can be used also to derive specific information for REST APIs, for example detecting the use of
query parameters not specified in the OpenAPI schema.
The papers which have addressed this issue such as~\cite{\paperfiftieth,\paperseventieth} transform the code of SUT to improve the fitness function during the search.

\textit{Mocking} is another addressed challenge by two of the papers~\cite{\papertwentysecond,\paperseventysixth} which is aimed at providing reliable response for the services that might not be accessible or down temporarily.
\textit{Instance Identification} is a problem in the context of micro-service testing that software testers face as they might not know which concrete instance of service is being invoked.
Vassiliou-Gioles has suggested adding a micro-service instance identification (IID) header field to HTTP requests and responses in order to make them more testable~\cite{\paperfiftyseventh,\paperfiftyeighth}.

How to measure the effectiveness of black-box test generators is a challenge addressed in~\cite{\papertwentyfourth,\paperthirtythird}, where new black-box criteria besides counting detected faults have been defined (i.e., \textit{Defining coverage criteria}).
This is particularly important when testing remote APIs for which code coverage metrics cannot be used.

%To answer this question, we investigated papers that have formulated their research goals in the form of research question. We found \PapersWithRQs papers which have written down their research questions which resulted in gathering \ExtractedRQs research questions. As the papers included a variety of contribution types with many different ideas, the collected research questions were highly diverse. However, we tried to find out if there are any addressed question which are common among them. 
%
%Table~\ref{tab:rq11} includes the most common challenges that are posed by the research questions of the papers. As it is shown, \textit{Measuring Code Coverage} and \textit{Compare with existing approaches} are the most frequent challenges by being addressed by 22 papers each. Measuring code coverage refers to calculating code coverage, whether line, branch, method \etc. Compare with existing approaches includes evaluating the effectiveness of the proposed approach with existing ones. This does not include comparing the new approach with random methods (\eg \cite{\paperfiftysixth}). \textit{Fault detection}, such as \texttt{5xx} server errors, is another common challenge posed by the research questions.

\begin{result}
	{\bf RQ11}: The most common addressed challenges include handling 
	resource
	and inter-parameter dependencies, as well as defining new automated oracles.
\end{result}

%----------------------------------------------------------
\subsection{\rqDOpenChal}

\begin{table}[!ht]
	\small
	\centering
	\caption{Common Open Challenges}
	\label{tab:rq12}
	\begin{tabular}{ l l p{0.6\textwidth} } 
\toprule 
Challenge & \# & Papers \\ 
\midrule 
Tool support & 25 & \cite{\papersecond,\paperfifth,\paperninth,\papereleventh,\papertwelfth,\paperseventeenth,\papertwentythird,\papertwentyfourth,\papertwentysixth,\paperthirtieth,\paperfortyfirst,\paperfortysecond,\paperfortyfourth,\paperfortysixth,\paperfiftysecond,\papersixtyfifth,\papersixtysixth,\papersixtyninth,\paperseventyfourth,\paperseventyseventh,\papereightyfirst,\papereightysixth,\paperninetysecond,\paperninetyseventh,\paperninetyeighth}\\ 

Having more REST case-studies & 13 & \cite{\paperthird,\paperseventh,\papereighth,\paperfourteenth,\paperseventeenth,\papertwentyfirst,\papertwentyfourth,\papertwentyfifth,\paperthirtyeighth,\paperfortyfifth,\paperseventyfifth,\paperseventyninth,\paperninetieth}\\ 

Security testing & 10 & \cite{\paperfirst,\paperseventeenth,\papertwentyfirst,\papertwentyeighth,\papertwentyninth,\paperthirtieth,\paperfiftysecond,\papersixtysixth,\paperseventyseventh,\papereightyeighth}\\ 

Resource handling & 4 & \cite{\paperthird,\papertwentyfifth,\paperfortyfourth,\paperfortyfifth}\\ 

Classifying test results & 4 & \cite{\paperfifth,\papersixth,\paperninth,\paperfiftysecond}\\ 

Handling external services & 4 & \cite{\papersixth,\paperfortieth,\papersixtyfifth,\papereightyfirst}\\ 

White-box heuristics & 4 & \cite{\paperseventh,\paperfiftieth,\paperseventieth,\papereightyfirst}\\ 

Database handling & 4 & \cite{\papertwentythird,\paperthirtyninth,\paperseventyfirst,\papereightyfirst}\\ 

Automated oracles & 4 & \cite{\papersixtieth,\papereightieth,\papereightythird,\paperninetyfourth}\\ 

Performance testing & 3 & \cite{\papertwentyninth,\papersixtyninth,\papereightysecond}\\ 

Non-functional testing & 2 & \cite{\papereleventh,\paperninetysecond}\\ 

Load testing & 1 & \cite{\papersecond}\\ 

Metamorphic testing & 1 & \cite{\papereleventh}\\ 

Underspecified schemas & 1 & \cite{\paperfortieth}\\ 

\bottomrule 
\end{tabular} 

\end{table}

Most of the papers we studied have mentioned some left objectives, to be addressed in future work.
Among the selected papers, there were \PapersWithOpenChallenges papers which have explicitly mentioned at least one open challenge for future work.
This is usually stated in the \textit{Conclusion} section of these articles, or in specific \textit{Future Work} section.
We collected all of these objectives, and identified the most common challenges among them, as shown in Table~\ref{tab:rq12}.
Note that, in most cases, a typical challenge is to design better techniques to obtain better code coverage and fault finding results.
Here, we rather focus on more specific challenges related to testing RESTful APIs.

Based on Table~\ref{tab:rq12},
the most common objective left as future work is \emph{Tool support} that is mentioned in \toolsupport papers.
Research in this domain is based on tool prototypes.
Due to the complexity of handling RESTful APIs, these tools require major engineering effort.
An early prototype can provide the base to experiments with some research ideas, which can already be of use and be publishable.
Work is then left to improve these tools to be able to be applicable  on more case studies, and to be more user-friendly for practitioners.
For example, in~\cite{\papertwentysixth} the authors proposed an approach that supports OpenAPI v2, and plan  to enabling its support for OpenAPI v3 as well.
Another example, supporting authentication is needed to enable testing endpoints which needs the client to be authenticated, which  was specified as needed future work in~\cite{\paperseventeenth}.
In these cases, not all techniques presented in the literature support authentication, or the full specs of the OpenAPI standard.
Still, even with partial support, it is possible to provide and evaluate novel techniques.

Being mentioned in \havingmorerestcasestudies papers, \textit{Having more REST case-studies} to apply the new approach to a higher number of REST APIs for empirical evaluation was the second most frequent open challenge.
This is a general issue related to \emph{Threats to External Validity}, which applies to most empirical studies in the software engineering research literature.
However, one peculiarity here is on the kind of systems used for experimentation.
On the one hand, APIs on the internet pose few issues, including difficulty in replicating studies (APIs can change at any time) and inability to collect code coverage metrics.
On the other hand, running APIs on   local machines for experimentation has non trivial setup costs, for example on how to setup databases and user info authentication.
It can take a significant amount of time to find and setup a large number of REST APIs for experimentation.
Furthermore, experiments on system test generation for Web APIs are time consuming, as each test case evaluation requires to execute HTTP calls over a network.
Given a fixed amount of time to run experiment (e.g., a machine with experiments left running for a week), this reduces the number of APIs that can be used for experimentation.

Many of the articles discussed in this paper present a new approach for generating test cases that can find faults, based on functional properties (e.g., API should not return a 500 HTTP status code, or a response not matching the constraints of the schema).
A common line of future work mentioned in these articles is to extend such work to consider other types of testing,
in particular \emph{Security testing} (\securitytesting papers),
as well as others such as \emph{Performance testing} and \emph{Load testing} (and other unspecified \emph{Non-functional testing}).

The other types of open challenges are  mentioned less often, in up to \handlingexternalservices articles.
For example, there are specific properties of REST APIs (e.g., the idempotency of HTTP verbs) that could be used as
\emph{Automated oracles} to be able to find more faults.
Related, it will also be important to properly analyze the test results (i.e., \emph{Classifying test results}) to be able to automatically check if the obtained responses represent actual faults, and classify their importance/criticality.

Regarding \textit{Handling external services},
it is possible for a RESTful API to rely on communications with other RESTful APIs.
Whether the testing mode is either black box or white box, dealing with external services makes testing these APIs very difficult.
In the black box mode, there would be no control of the external services.
Interactions with these external services would be based on both their implementation and current status.
As a result, the chance of being flaky increases.
Handling external services is a challenge in white-box mode as well.
Even if a developer had complete control over all of those services, it might be difficult to set up and run numerous separate web services (each of which can utilize its own database) for usage by the SUT during testing.

Another issue is when dealing with \emph{Database handling}.
Databases are very common in RESTful APIs, and the interactions of the APIs with these databases do impact the level of code coverage and fault finding the fuzzers can reach.
Some basic techniques have been presented to handle SQL databases, but more needs to be done, especially to handle other kinds of databases, like NoSQL databases such as MongoDB.

When the content of databases cannot be analyzed (e.g., in black-box testing), it is important to detect relations between operations, such as the need to create a resource with a \texttt{POST} request first before being able to test its \texttt{GET} endpoint.
Several articles have addressed this issue (e.g., recall Section~\ref{sec:AdresChal}).
Still, a major issue regarding \emph{Resource handling} is how to deal with schemas that do not follow the guidelines of REST (e.g., resources not structured hierarchically), as it is much harder there to infer resource relations among the different endpoints.

Based on our review, only one tool supports automated white-box testing of RESTful APIs (i.e., \evo).
As the achieved code coverage still needs to be improved, several code-level issues have been identified and categorized, which will need to be addressed to achieve higher code coverage, e.g., with new \emph{White-box heuristics}.
One possible venue where white-box heuristics can be useful is to deal with \emph{Underspecified schemas}, i.e., when there are constraints on the data and operations of the API, but those are not specified in the schema.

\begin{result}
	{\bf RQ12}:
	We have identified a variety of open challenges discussed in the analyzed articles.
	There is still a lot of research work that is needed to be carried out in this domain.
	%Among them, having more REST case studies was the most significant one by being mentioned in \havingmorerestcasestudies papers.
\end{result}

%%%%%%%%%%%%%%%%%%%%%%%%%%%%%%%%%%%%%%%%%%%%%%%%%%%%%%%%%%%%%%%%%%%%%%%%%%%%

\section{Threats to Validity}
\label{sec:threats}

As for any survey, there is the validity threat that some important and relevant articles have been missed from our analysis.
We used the most popular search databases to find all relevant articles.
%, and followed both a forward and backward snowballing procedure to find any other missing articles.
As the results reply on search engine provide by each database and some relevant paper might not be published in the selected databases, we additionally performed
a forward and backward snowballing procedure to find any other missing articles.
In addition, paper selection was 
%However, as this procedure was 
done manually, then human mistakes are possible.
To reduce such a risk, 
%two 
three authors were involved in the procedures in order to reach a final agreement by all, e.g.,
%Furthermore, the decision of which paper to exclude was done manually as well.
%
only papers that all of us the three authors agreed on to exclude were excluded. 
However, there is always the possibility that other researchers might have come to some different selections.
%We are confident that we have included the most relevant articles, but we cannot guarantee that a few ones have not been missed. 

Extracting data from the articles required manual effort and expertise, and it might be prone to human mistakes.
To reduce such validity threat, each selected paper was checked by at least two of us authors.
We are the authors of  \NumberOfPapersByUs articles out of the \totalPapers in our survey.
Although we are confident that we were able to properly analyze our own work, there is always the risk that we might have misunderstood the text of some articles written by other researchers.

%%%%%%%%%%%%%%%%%%%%%%%%%%%%%%%%%%%%%%%%%%%%%%%%%%%%%%%%%%%%%%%%%%%%%%%%%%%%

\section{Conclusion}
\label{sec:conclusions}

In this survey, we have collected and analyzed \totalPapers articles on the topic of testing RESTful APIs.
For this analysis, we can see that there has been an exponential increase in interest on this topic in the research community, starting from 2017 (Section~\ref{sec:survey}).
Many different techniques have been evaluated, including both black-box and white-box techniques (Section~\ref{sec:approaches}).
Furthermore, besides scientific articles, several prototypes have been released as open-source projects, with empirical investigation carried out on many real-word APIs, finding real faults in them.
This shows potential usefulness of this line of research for practitioners in industry (Section~\ref{sec:tools}).
Different scientific challenges have been addressed, while others still need to be solved (Section~\ref{sec:challenges}).

RESTful APIs are widely used in industry.
Research work on this topic has strong potential to have significant impact on industrial practice.

This survey provides a detailed snapshot of the current state-of-the-art in the research literature on testing RESTful APIs.
This is a growing field, where this survey can provide a useful starting point to drive new research directions on this important topic.

%%%%%%%%%%%%%%%%%%%%%%%%%%%%%%%%%%%%%%%%%%%%%%%%%%%%%%%%%%%%%%%%%%%%%%%%%%%%
\section*{Acknowledgments}
This work is funded by the European Research Council (ERC) under the European Union’s Horizon 2020 research and innovation programme (EAST project, grant agreement No. 864972).

%https://arxiv.org/help/submit_tex#latex
% USE generated bbl
%\bibliographystyle{acm}
%\bibliography{papers}

\end{document}